\documentclass[%
 reprint,
 amsmath,amssymb,
 aps,
]{revtex4-2}

\usepackage{graphicx}% Include figure files
\usepackage{dcolumn}% Align table columns on decimal point
\usepackage{bm}% bold math

\begin{document}

\preprint{APS/123-QED}

\title{Spin waves propagating through a stripe magnetic domain structure and \\their applications to reservoir computing}% Force line breaks with \\


\author{Ryosho Nakane$^1$}
\author{Akira Hirose$^1$}
\author{Gouhei Tanaka$^{1, 2}$}
\affiliation{%
$^1$Department of Electronic Engineering and Information Systems, Graduate School of Engineering, \\ The University of Tokyo, 7-3-1 Hongo Bunkyo-ku, Tokyo 113-8656, Japan \\
$^2$International Research Center for Neurointelligence (IRCN), The University of Tokyo, 7-3-1 Hongo Bunkyo-ku, Tokyo 113-0033, Japan
}%


%\author{}
%\affiliation{%
%\affiliation{

%\affiliation{%


\date{\today}% It is always \today, today,

\begin{abstract}

Spin waves propagating through a stripe domain structure and reservoir computing with their spin dynamics have been numerically studied with focusing on the relation between physical phenomena and computing capabilities. 
Our system utilizes a spin-wave-based device that has a continuous magnetic garnet film and 1-input/72-output electrodes on the top.  To control spatially-distributed spin dynamics, a stripe magnetic domain structure and amplitude-modulated triangular input waves were used. 
The spatially-arranged electrodes detected spin vector outputs with various nonlinear characteristics that were leveraged for reservoir computing. By moderately suppressing nonlinear phenomena, our system achieves 100$\%$ prediction accuracy in temporal exclusive-OR (XOR) problems with a delay step up to 5. At the same time, it shows perfect inference in delay tasks with a delay step more than 7 and its memory capacity has a maximum value of 21. This study demonstrated that our spin-wave-based reservoir computing has a high potential for edge-computing applications and also can offer a rich opportunity for further understanding of the underlying nonlinear physics.

%\begin{description}
%\end{description}
\end{abstract}

\maketitle

%\tableofcontents

\section{Introduction}

Reservoir computing is a computational framework which is originally based on recurrent neural networks \cite{jaeger2001echo, maass2002real}. It is realized with a system having a reservoir part and a readout part. In this computational framework, the role of the reservoir part is to nonlinearly transform time-series input data to high-dimensional spatiotemporal signals, which allows us to optimize the readout part with linear regression while the reservoir part is unchanged. Owing to its unique feature, reservoir computing models require much less training cost than deep neural networks. Thus, it is promising for machine-learning-based edge computing \cite{shi2016edge, abbas2017mobile} that can perform energy-efficient information processing of a time-series data obtained from mobile devices and sensors. Recently, it has been demonstrated that reservoir computing systems can be realized with reservoirs based on physical phenomena. Such reservoir computing, called physical reservoir computing \cite{tanaka2019recent}, can create machine-learning electronic devices that are directly mounted onto terminal devices, leading to revolutionary technologies in next-generation Internet-of-Things (IoT) era. The following examples are some of the typical systems and devices used for physical reservoir computing: optelectronic systems \cite{larger2012photonic, paquot2012optoelectronic}, optical systems \cite{duport2012all, brunner2013parallel}, an electronic circuit \cite{appeltant2011information}, memristive resistance networks \cite{du2017reservoir, moon2019temporal}, a soft material \cite{nakajima2015information}, water in a bucket \cite{fernando2003pattern}, a silicon beam \cite{dion2018reservoir}, a ferroelectric metal-oxide-semiconductor field-effect transistor \cite{nako2020proposal}, spin torque oscillators \cite{torrejon2017neuromorphic, kanao2019reservoir, yamaguchi2020periodic}, a magnetic tunnel junction \cite{furuta2018macromagnetic}, nanomagnets \cite{nomura2019reservoir}, and spin waves \cite{nakane2018reservoir, nakane2019spin}. Although the relation between physical properties and computation capabilities in these physical systems has not been fully understood, physical reservoir computing can be performed well when a physical system possesses rich dynamics including high-dimensionality, nonlinearlity, input-history dependency, and fading memory property (echo state property) \cite{tanaka2019recent}. In particular, high capabilities in extremely-efficient information processing are expected for excitable continuous medium reservoirs utilizing propagation of waves triggered by stimulation inputs \cite{fernando2003pattern}, without internal wiring. In these reservoirs, a high dimensionality can be realized by large numbers of spatially-arranged inputs/detectors for input/output signals. To take advantage of this characteristic, it is primarily important to excite, control, and detect waves to have rich dynamics.
\begin{figure*}[bt]
\includegraphics[width=\linewidth]{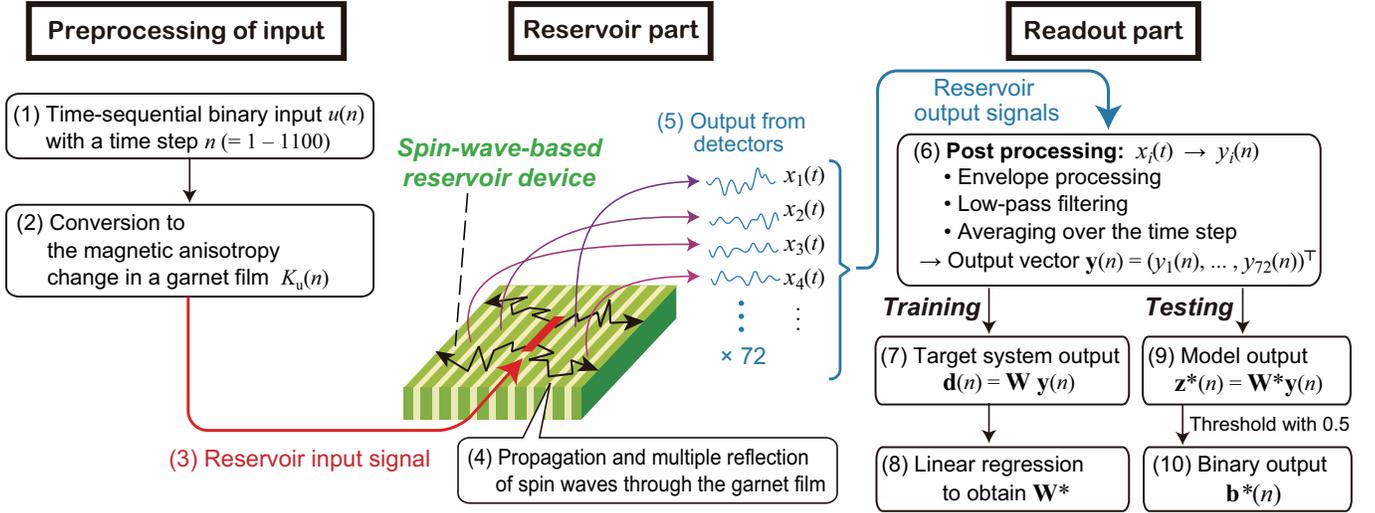}
\caption{\label{FIG_1}Spin-wave-based reservoir computing system that is composed of an input preprocessing, a reservoir part, and a readout part. The computing is executed following the order of (1) – (8) in the traning, whereas it is executed following the order of (3) – (6) and then (9) – (10) in the testing. In a spin-wave-based reservoir device in the reservoir part, a red area is the input exciter, a zebra pattern with pale and deep greens represents a stripe magnetic domain structure, and bended black arrows schematically illustrate the expected propagation of spin waves with multiple reflections by the magnetic domains. Blue curves schematically illustrate time-series output waveforms obtained at 72 detectors in the device.}
\end{figure*}

In our recent papers \cite{nakane2018reservoir, nakane2019spin}, a spin-wave-based reservoir computing device has been proposed as an $on$-$chip$ excitable continuous medium reservoir and its application to machine-learning computation has been demonstrated. The spin-wave-based reservoir computing device is fairly attractive for applications in edge domains, since it can be feasibly realized on a chip with electrical wires just for input/output electrodes and its power consumption is expected to be low because of the signal transmission using spin waves \cite{mahmoud2020introduction}. Moreover, since spin waves originate from the dynamical change in the spin distribution that is one of the electronic properties, highly-torelant computing is expected. Toward practical applications, steady progress in computing capabilities can be made by starting with simple benchmark tasks: temporal exclusive-OR (XOR) problems and delay tasks for evaluating memory capacity. For the reservoir system, the former task requires both nonlinearity and short-term memory, whereas the latter task requires only short-term memory. Since these characteristics play crucial roles in reservoir computing, the achievements of high computing capabilities for these tasks clearly show high potential of the system. Furthermore, a comprehensive study on spin-wave-based reservoir computing can provide a deep insight into the relation between physical phenomena and computing capabilities, which is one main interest in the field of physical reservoir computing.

In this paper, we numerically study reservoir computing using a spin-wave-based reservoir device. To efficiently realize spatially-distributed rich dynamics of spin waves with nonlinear phenomena, including nonlinear propagation and multiple reflections, amplitude-modulated waves representing a bit sequence are used for input signals and a stripe magnetic domain structure is introduced in a continuous magnetic garnet film where spin waves propagate. It appears that various waveforms are obtained at different positions in the magnetic garnet film by controlling nonlinear phenomena with input and material parameters. Then, using the resultant waveforms under various parameter conditions, benchmark tasks are solved to reveal what features of spin waves are effective for high-capability reservoir computing. In temporal XOR problems with various interval time steps, it is found that the prediction accuracy strongly depends on the above two parameters. Under moderate suppression of nonlinear phenomena in spin waves, 100$\%$ prediction accuracy is achieved for relatively-large delay steps up to 5. At the same time, the memory capacity estimated using delay tasks with various delay time steps has a maximum value of 21. Finally, the relation between physical phenomena and computing capabilities is discussed.

\section{Reservoir Computing System}
Spin-wave-based reservoir computing was performed with simulator-calculated spin waves. Figure \ref{FIG_1} shows a schematic of our reservoir computing system that consists of a signal preprocessing part, a spin-wave-based reservoir part, and a readout part. In this study, the discrete time step is expressed by $n (= 1, 2, \cdots, 1100)$ and the input data is expressed by $u(n)$ that has a bit value (0 or 1) changing randomly with $n$.

First, in the preprocessing part, $u(n)$ is transformed to a reservoir input signal that is the change in the uniaxial magnetic anisotropy $K_\textup{u}(n)$ of the magnetic garnet film in the reservoir device. When a time-series reservoir input signal $K_\textup{u}(n)$ is fed into the input exciter of the magnetic garnet film (the red line), spin waves are excited and then propagated. Since the magnetic garnet film has a stripe domain structure illustrated by pale and very-pale green stripes, as shown later, it is expected that spin waves reflected multiple times interfere with each other, as illustrated by black bended arrows.  Reservoir output signals $x_i(t)$ $(i = 1, 2, \cdots, 72)$ in response to the reservoir input signal are obtained by detecting spin waves at the $i$th detector on the magnetic garnet film, where $t$ represents continuous time. To use steady response, the output signals in the first 100 time steps (corresponding to the first 100 bits) are discarded in the readout processing. 

In the readout part, $x_i(t)$ are converted to $y_i(n)$ through some signal processing: an envelope processing, a low-pass filtering, and an averaging over the time step range. Then, an optimization of the readout part is performed based on a linear regression \cite{lukovsevivcius2009reservoir}. By collecting $y_i(n)$ for $i = 1, 2, \dots, N_y$ with $N_y \le 72$, a reservoir output vector $\mathbf{y}(n)$ is given by 
\begin{eqnarray}
\mathbf{y}(n) &=& (y_1(n), \dots, y_{N_y}(n))^\mathsf{T} \in \mathbb{R}^{N_y}.
\end{eqnarray}
The system output at the $k$th node is expressed by $z_k(n)$ and then a system output vector $\mathbf{z}(n)$ is given by
\begin{eqnarray}
\mathbf{z}(n) &=& (z_1(n), \dots, z_{N_z}(n))^\mathsf{T} \in \mathbb{R}^{N_z}.
\end{eqnarray}
The system output is expressed by 
\begin{eqnarray}
\mathbf{z}(n) &=& \mathbf{W}^\textup{out}\mathbf{y}(n),
\end{eqnarray}
where $\mathbf{W}^\textup{out}$ is an output weight matrix: 
\begin{eqnarray}
\mathbf{W}^\textup{out} &=& (w^\textup{out}_{ki}) \in \mathbb{R}^{N_z \times N_y}.
\end{eqnarray}
 After the reservoir outputs $\mathbf{y}(n)$ from $n$ = 101 to 1100 are divided into two parts, 500 steps in the first part are used for training, whereas 474 steps in the second part are used for testing. In the training, the target signal $d_k(n)$ for the $k$th system output in each task is generated using $u(n)$ and then the target of the system output vector $\mathbf{d}(n)$ is given by
\begin{eqnarray}
\mathbf{d}(n) &=& (d_1(n), \dots, d_{N_z}(n))^\mathsf{T} \in \mathbb{R}^{N_z}.
\end{eqnarray}
In the traning, to determine the optimum weight matrix $\mathbf{W}^*$, $\mathbf{Y}$ and $\mathbf{D}$ matricies are given by column-wise collections of $\mathbf{y}(n)$ and $\mathbf{d}(n)$, respectively, as follows:
\begin{eqnarray}
\mathbf{Y} &=& [\mathbf{y}(101), \dots, \mathbf{y}(500)] \in \mathbb{R}^{N_y \times 500}\\
\mathbf{D} &=& [\mathbf{d}(101), \dots, \mathbf{d}(500)] \in \mathbb{R}^{N_z \times 500}.
\end{eqnarray}
Then, $\mathbf{W}^*$ is calculated by using the pseudo-inverse matrix $\mathbf{Y}^{\dagger}$:
\begin{eqnarray}
\mathbf{W}^* &=& \mathbf{D}\mathbf{Y}^{\dagger}.
\end{eqnarray}

In the testing, a model output signal vector $\mathbf{z}^*(n)$ is calculated using $\mathbf{W}^*$ and $\mathbf{y}(n)$ as follows:
\begin{eqnarray}
\mathbf{z}^*(n) &=& \mathbf{W}^*\mathbf{y}(n).
\end{eqnarray}
Then, a binary output signal $b_k^*(n)$ at the $k$th node is calculated from $z_k^*(n)$ with a threshold of 0.5 as follows:
\[
  b_k^*(n) = \left\{ \begin{array}{ll}
    1 & (z_k^*(n) \geq 0.5) \\
    0 & (z_k^*(n) < 0.5).
  \end{array} \right.
\]

For evaluating the computing capability with the system output at the $k$th node, the prediction accuracy of a computing task in the testing is calculated using the normalized Hamming distance as follows:
\begin{eqnarray}
\textup{ACC} = 1 \,–\, \sum_{n}\frac{h(d_k(n), b_k^*(n))}{L},
\end{eqnarray}
where $h(\cdot, \cdot)$ denotes the Hamming distance and $L$ = 474 is the total time step length used in the testing. 

\begin{figure}[b]
\includegraphics[width=7.0 cm]{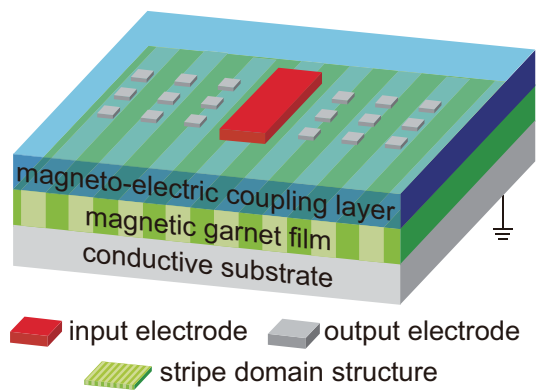}
\caption{\label{FIG_2}Schematic illustration of a spin-wave-based reservoir computing device that is composed of (from the bottom to top) a conductive substrate, a magnetic garnet film, a magneto-electric coupling layer, and input (a red cuboid)/output (gray cylinders) electrodes. The stripes in the magnetic garnet film denote a stripe magnetic domain structure. In the operation of an actual device, spin waves in the magnetic garnet film are excited by an input voltage, they propagate through the magnetic garnet film, and their dynamics beneath the output electrodes are detected by output voltages.}
\end{figure}

\section{Spin-wave-based Reservoir Computing Device}
\subsection{Device Structure}
Figure \ref{FIG_2} shows a schematic device structure that is composed of (from the bottom to top) a conductive substrate, a magnetic garnet film, a magneto-electric (ME) coupling layer, and input (a red cuboid)/output (gray cuboids) electrodes.  It is basically the same as that in our previous paper \cite{nakane2018reservoir} and the main differences are the shape of the input electrode, the arrangement of the output electrodes, and the stripe magnetic domain structure. The magnetic garnet film/ME coupling bilayer structure is typically called synthetic multiferroic structure that converts from magnetic to electrical properties, and vice versa. In the operation of an actual device, spin waves in the magnetic garnet film are excited and detected by the voltages of the input and output electrodes, respectively, through the function of the synthetic multiferroic structure, such as voltage controlled magnetic anisotropy (VCMA) \cite{rana2019towards}. 

In numerical experiments in this study, since spin dynamics in the magnetic garnet film is simulated, input and output are expressed by the change in the properties of the magnetic garnet film near the surface, as a typical ME coupling in synthetic multiferroic structures. The input is expressed by the change in the magnetic anisotropy near the surface.  The output is expressed by the out-of-plane component of the averaged spins over a small volume beneath each output electrode since the output voltage is expected to be proportional to the out-of-plane component of spins near the surface.

\subsection{Key Ideas for High Performance in Reservoir Computing\label{Idea}}
The followings are our consideration on what physical phenomena are specifically needed in our spin-wave-based reservoir computing device:
\begin{itemize}
\item High-dimensionality: Different waveforms are obtained at different positions in space.
\item Nonlinearlity: Nonlinear phenomena, such as nonlinear propagation and interference, occur in response to a time-series input signal.
\item Input-history dependency and fading memory property: Spin waves excited by an input signal that corresponds to one bit infromation causes fluctuations at the same position for a certain time length.
\end{itemize}
In physical reservoir computing, the reservoir part appropriately designed in advance is unchanged. Thus, the input transformation from a time-series data to a time-series physical quantities and the configuration of a reservoir device are essential for the realization of rich physical dynamics. Our key ideas for these two building blocks are described below.  

Regarding input signals, our idea is to use amplitude-modulated waves corresponding to bit sequence data. This is because transient nonlinear responses to input signals are effective for reservoir computing \cite{appeltant2011information} and such responses excited by forced oscillations can be reliably measured in the signal envelope, as in spin torque oscillators \cite{torrejon2017neuromorphic}. When the input $u(n)$ is transformed to a reservoir input signal, the time period of the modulation signal is an important parameter, since it is expected that nonlinear transient phenomena strongly depend on it. Thus, this study examines how spin dynamics changes depending on this time period.

Regarding the configuration of a reservoir device, our idea is to realize unstable spin waves in a stripe domain structure that is frequently obtained in magnetic garnet films under near-zero external DC magnetic field $H^\textup{EX}$ \cite{hubert2008magnetic}. Here, ``unstable'' is used to indicate a situation that long-range dipole interactions are not negligible due to large polar angles in spin precessions and the precession axis for each spin dynamically changes. From previous papers by other groups \cite{stancil2009spin}, unstable spin waves can lead to rich dynamics, including period-doubling bifurcations and chaotic behavior, when a high-power radio-frequency (RF) input is used. Instead of this method, when a stripe domain structure with a small $H^\textup{EX}$ is used, the following features are expected: Unstable spin waves result in space-varying spin dynamics, they show nonlinear phenomena, and they are split and reflected multiple times in the presence of domains (and domain boundaries). These can lead to the aforementioned physical characteristics needed for reservoir computing.  Since the properties of the magnetic garnet film are important parameters, this study examines how spin dynamics changes depending on the damping factor at the boundaries of the garnet film, as will be specifically explained in Sec. \ref{Procedure}. It is noteworthy that a small $H^\textup{EX}$ can be realized by an available technology for on-chip device \cite{craik1978bias}, which is advantageous for feasible implementation.

\subsection{\label{Procedure}Simulation Procedure and Material Parameters}
\begin{figure}[tb]
\includegraphics[width=\linewidth]{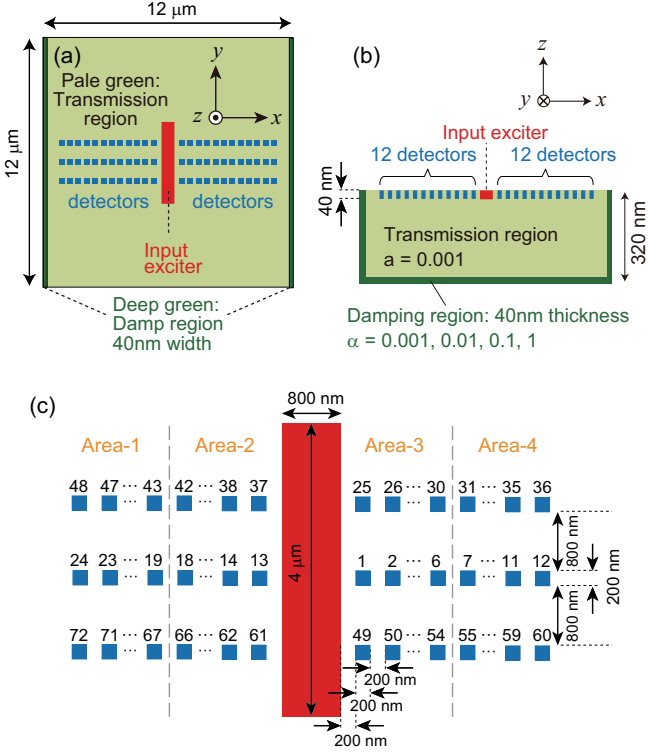}
\caption{\label{FIG_3}(a) Top and (b) cross-sectional views of the spin-wave-based reservoir computing device utilizing a magnetic garnet film, where a Cartesian coordinate is defined in each figure, pale-green and deep-green regions are the transmission and damping regions of spin waves in the garnet film, respectively, and red and blue regions are the input exciter and detectors, respectively. The damping constant $\alpha_0$ in the transmission region is 0.001, whereas the damping constant $\alpha$ in the damping regions is varied ($\alpha$ = 0.001, 0.01, 0.1, and 1) to examine the change in spin dynamics, as described in \ref{Idea}. The $x$-$y$ plane and thickness along the $z$ axis of the magnetic garnet film is 12 $\times$ 12 $\mu\textup{m}^2$ and 320 nm, respectively. The depths of the input exciter and detectors are 40 nm from the surface, as shown in (b). (c) Detailed top-view structure of the input exciter and detectors on the top of the garnet film, where detectors with 200 $\times$ 200 $\textup{nm}^2$ are arranged with 200 nm and 800 nm in pitch along the $x$ and $y$ axes, respectively, and the number is defined for each detector. All the detectors are divited into 4 groups Area-1, -2, -3 and -4, each of which has 18 detectors.}
\end{figure}

Figures \ref{FIG_3}(a) and (b) show top-schematic and cross-sectional-schematic views of a magnetic garnet film used for the simulation, respectively, where red and blue regions are the input and output regions, respectively, and a Cartesian coordinate system is defined in each figure. Hereafter, the input and output region are called the input exciter and detector, respectively. The $x$-$y$ plane has an area of 12 $\times$ 12 $\mu \textup{m}^2$ and a thickness of 320 nm along the $z$ direction, where pale-green and deep-green regions are the transmission and damping regions of spin waves, respectively, and the transmission region includes the input exciter and detectors. The upper and lower boundaries are connected for reducing the simulation time, which is expected to have no influence on detected spin waves since the detectors are arranged far from these boundaries. 
Figure \ref{FIG_3}(c) shows the detailed top-view arrangement of the input exciter (a red rectangle) and detectors (blue rectangles) in the magnetic garnet film, respectively. The input exciter has an area of 400 nm $\times$ 4 $\mu\textup{m}$ centered at the origin of the $x$-$y$ plane and has a depth of 40 nm along the $z$ axis (Fig. \ref{FIG_3}(b)). The detectors have an area of 200 nm $\times$ 200 nm and a depth of 40 nm along the $z$ axis (Fig. \ref{FIG_3}(b)), they are arranged at a 400 nm pitch interval along the $x$ axis and at a 1000 nm pitch interval along the $y$ axis, respectively, and the detector number is defined in Fig. \ref{FIG_3}(c). The detectors are divided into 4 groups Area-1, -2, -3, and -4  by the broken-gray boundaries along the $y$ axis and each group has 18 detectors. The damping regions at the left and right boundaries are expected to fully damp the spin waves propagating from the input region at the center, i.e., there is no reflection of spin waves from the boundaries to the center. The reflection of spin waves will be verified later.

We used a micromagnetic simulator Mumax3 \cite{vansteenkiste2014design} for numerical experiments on spin dynamics in the magnetic garnet film in Figs. \ref{FIG_3}(a) and (b). The simulation and material parameters are as follows: The mesh unit defined by the $xyz$ axes is cubic with 40 nm on one side, a spin (magnetic moment) is located at every mesh corner, the simulation temperature is 0 K, the saturation magnetization $M_\textup{s}$ is 190 kA/m, the stiffness constant $A_\textup{EX}$ is  $3.7 \times10^{-12}$ J/m, the uniaxial magnetic anisotropy $K_\textup{u}$ along the $z$ axis is $K_\textup{u}^\textup{H}$ = 5 $\textup{kJ}/\textup{m}^3$ in the damping region and the transmission region except the input exciter, and the damping constant $\alpha_0$ is 0.001 in the transmission region. The $M_\textup{s}$ value is that of a bulk $\textup{Y}_3\textup{Fe}_5\textup{O}_{12}$ (YIG)  at a few K \cite{anderson1964molecular}, the $A_\textup{EX}$ value is identical with the value for a liquid-sphase epitaxy (LPE) grown YIG film on a $\textup{Gd}_3\textup{Ga}_5\textup{O}_{12}$ (GGG) substrate \cite{klingler2014measurements}, and the $K_\textup{u}^\textup{H}$ is similar to the values for pulsed laser depsition (PLD) grown YIG films on GGG substrates \cite{manuilov2009pulsed}. In the input exciter, $K_\textup{u}$ along the $z$ axis was changed in proportion to an input signal, as will be explained in Sec. \ref{Input}. To study fundamental properties and reservoir computing with various types of spin dynamics, $\alpha$ in the damping region was varied (= 0.001, 0.01, 0.1, and 1). During the simulation, a constant external magnetic field $H^\textup{EX}$ was applied along the +$y$ direction. The time step in the simulation was $1 \times 10^{-12}$ s, whereas averaged spin values for a specific detector were recorded at every $5 \times 10^{-11}$ s.  Hereafter, the $x$, $y$, and $z$ components of the normalized spin are expressed by $s_x$, $s_y$, and $s_z$, respectively. 

It should be noted that the $\alpha_0$ value 0.001 in the transmission region is larger by one order of magnitude than that experimentally-estimated values in epitaxial YIG films with high quality \cite{dubs2020low,pirro2014spin,ding2020sputtering,schmidt2020ultra}. The reasons are simply to reduce simulation time and to eliminate the influence of spin waves reflected back from the left and right boundaries by intentionally reducing the amplitudes of spin waves while propagating (the latter will be discussed in \ref{XOR}). Thus, since this $\alpha_0$ value just leads to smaller decay lengths of propagating spin waves, consistent results with those in this study will be obtained when simulation is performed using a YIG film having a larger $x$-$y$ plane and a smaller $\alpha_0$ value.

\subsection{Magnetic Domain Structure}
\begin{figure}[b]
\includegraphics[width=6 cm]{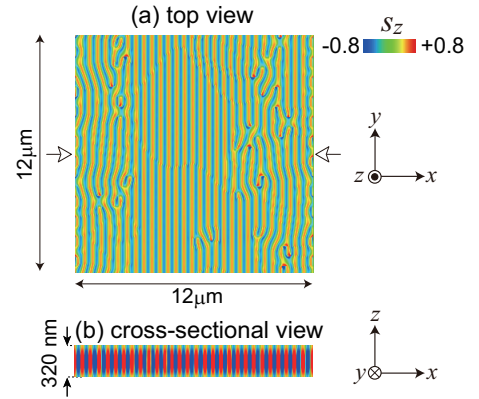}
\caption{\label{FIG_4}(a) Top and (b) cross-sectional views of the distributions of $s_z$ in the magnetic garnet film after the relaxation under a magnetic field $\mu_0 H^\textup{EX}$ = 0.005 T along the $+y $ direction, where the magnitude is defined by a color bar. The cross-sectional view is along the $x$ axis between two open arrows and it is extended along the $z$ axis for visibility.}
\end{figure}

The initial magnetic domain structure for spin dynamics simulation was formed using the following procedure: First $\mu_0 H^\textup{EX}$ = 0.05 T was applied, then it was reduced to 0.01 T in step of 0.01 T, and finally it was reduced to 0.005 T in step of 0.001 T. At each step, the spin distribution was relaxed. Figures \ref{FIG_4}(a) and (b)  show top and cross-sectional views of the distributions of $s_z$ in the magnetic garnet film after the final step, respectively, where the magnitude is defined by a color bar and the cross-sectional view is extended along the $z$ axis for visibility. The magnetic domain has a checkered pattern composed of red and blue lines along the $y$ axis, i.e., a stripe domain structure with two domains saturating along the $\pm$ $z$ directions. The deviation from the complete stripe structure increases as the $x$ position approaches the left and right boundaries. This feature is preferable for reservoir computing because it can lead to various waveforms at the detectors.

\subsection{\label{Input} Reservoir Input Signals}
\begin{figure}[tb]
\includegraphics[width=6.5 cm]{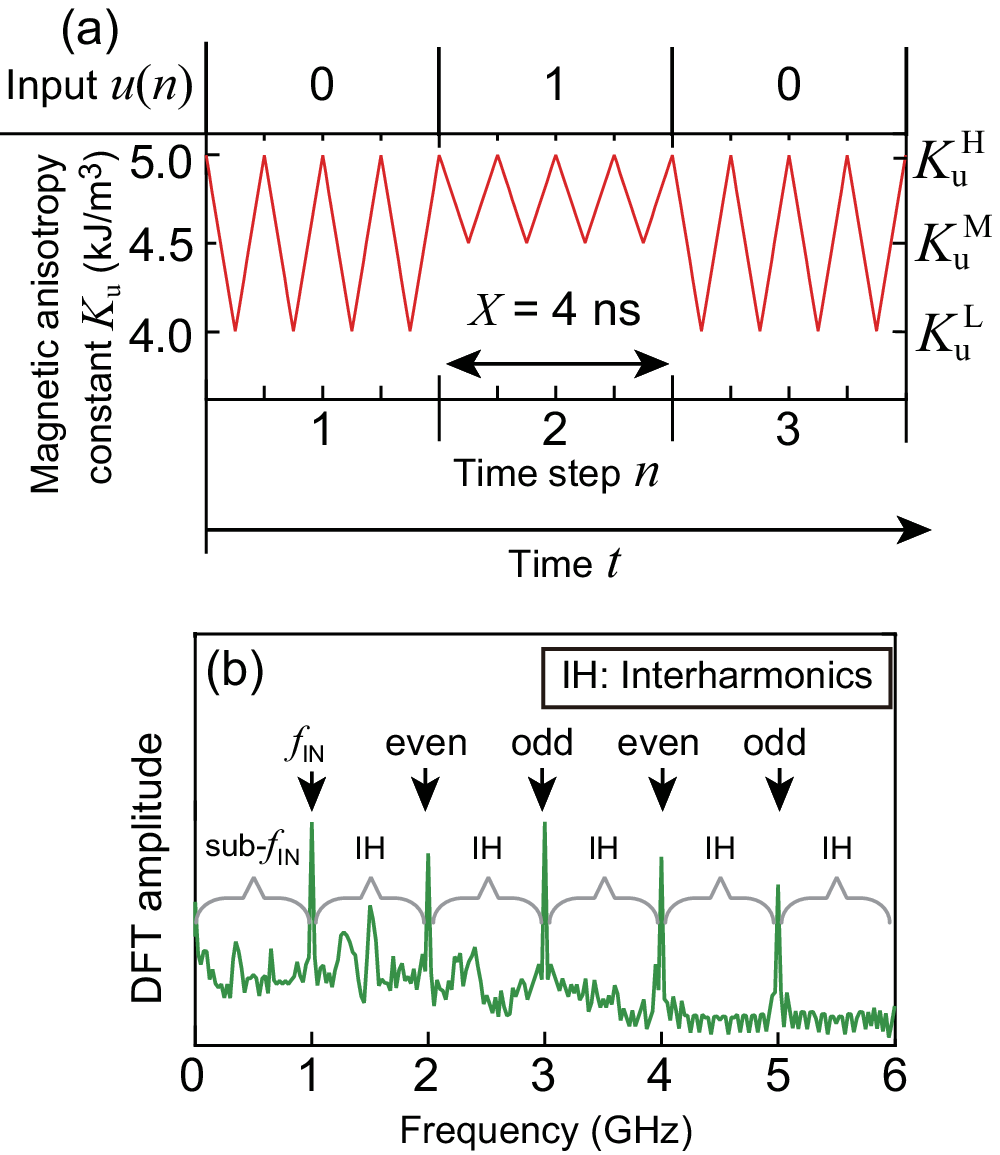}
\caption{\label{FIG_5} (a) Example of a reservoir input signal $K_\textup{u}(n)$ with the time step length $X$ = 4 ns: 0 and 1 in $u(n)$ are expressed by $\Delta K_\textup{u, 0}$ = $K_\textup{u}^\textup{H} - K_\textup{u}^\textup{L}$ and $\Delta K_\textup{u, 1}$ = $K_\textup{u}^\textup{H} - K_\textup{u}^\textup{M}$, respectively, with a use of $K_\textup{u}^\textup{H}$ as the baseline, where $K_\textup{u}^\textup{H}$, $K_\textup{u}^\textup{M}$, and $K_\textup{u}^\textup{L}$ are 5.0, 4.5, and 4.0 $\textup{kJ}/\textup{m}^3$, respectively. To analyze the effect of $X$ on the properties of spin waves as well as on the capability of the reservoir computing, reservoir input signals with various $X$ values were prepared; $X$ = 1, 2, 4, 6, and 8 ns. The repetition number of triangular pulses for one bit can be calculated such that $X$ ns is divided by 1 ns. (b) Schematic illustration of a DFT spectrum, where $f_\textup{IN}$, odd, and even denote the pulse repetion frequency (1 GHz), the odd-number harmonic frequencies (3 and 5 GHz), and the even-number harmonic frequencies (2 and 4 GHz), respectively, and sub-$f_\textup{IN}$ and IH denote interhamonics below $f_\textup{IN}$ and interharmonics between the interger-number harmonic frequencies, respectively.}
\end{figure}

In the preprocessing part in Fig. \ref{FIG_1}, $u(n)$ was transformed to the reservoir input signal $K_\textup{u}(n)$ as follows. First, the time lengths of a triangular pulse and the time step length $X$ were set at 1 ns and a multiple of 1 ns, respectively. As described in Sec. \ref{Idea}, $X$ is the time period of the modulation signal, which is an important parameter for controlling spin dynamics. To analyze the effect of $X$ on spin dynamics as well as on the capabilities in reservoir computing, responses to various $X$ values were examined: $X$ = 1, 2, 4, 6, and 8 ns. The repetition number of triangular pulses for one bit can be calculated such that $X$ ns is divided by 1 ns. For example, when $X$ = 4 ns, the repetition number is 4. Next, to change the amplitude of $\Delta K_\textup{u}$ following the input data, $K_\textup{u}^\textup{M}$ = 4.5 and $K_\textup{u}^\textup{L}$ = 4 $\textup{kJ}/\textup{m}^3$ were prepared with a use of the maximum $K_\textup{u}^\textup{H}$ = 5.0 $\textup{kJ}/\textup{m}^3$ as the baseline. As an example, Fig. \ref{FIG_5}(a) shows a reservoir input signal with $X$ = 4 ns, in which  0 and 1 in $u(n)$ are expressed by $\Delta K_\textup{u, 0}$ = $K_\textup{u}^\textup{H} - K_\textup{u}^\textup{L}$ and $\Delta K_\textup{u, 1}$ = $K_\textup{u}^\textup{H} - K_\textup{u}^\textup{M}$, respectively.  Finally, time-series reservoir input signals with various $X$ were prepared using the same $u(n)$ for $n = 1, \dots, 1100$. 

\subsection{Analysis of Spin Dynamics}
To clarify the relation between physical phenomena and computing capabilities, it is necessary to analyze how spin dynamics are changed by the values of the two parameters $\alpha$ and $X$. Hence, reservoir output waveforms at the detectors were observed and they were also characterized by discrete Fourier transform (DFT) spectra that were calculated using output waveforms in the $n$ range from 101 to 1100. In input signals, since a triangular pulse with a period of 1 ns is repeated, namely, a triangular wave with 1 GHz is used as a carrier wave, a DFT spectrum always has a component at 1 GHz. Hereafter, the input signal frequency 1 GHz is denoted by $f_\textup{IN}$. In the analysis, a DFT spectrum is divided into four parts with different names, following each feature \cite{testa2007interharmonics}, as shown in Fig. \ref{FIG_5}(b):
\begin{itemize}
\item Odd-number harmonics: waves characterized by DFT components at 1, 3, and 5 GHz that are the fundamental frequencies for a triangular pulse. 
\item Even-number harmonics: waves characterized by DFT components at 2 and 4 GHz.
\item Interharmonics: waves characterized by DFT components between the odd- and even-number harmonic frequencies, including the fractional-order harmonics at 1.5, 2.5, 3.5, and 4.5 GHz. 
\item Sub-$f_\textup{IN}$: waves characterized by DFT components below 1 GHz, including sub-harmonics, and super-subharmonics. In a broad definition, this part is included in interharmonics.
\end{itemize}
Since triangular pulses were used in input signals, the odd-number harmonics are linear responses, whereas the others are nonlinear responses due to nonlinear phenomena.  Hereafter, the odd-number harmonics and even-number harmonics are sometimes collectively called the integer-number harmonics. To obtain accurate components at the integer-number and fractional-number harmonic frequencies in DFT spectra, the window function in the calculation was rectangle and the time length was the maximum common multiples of the time periods of these harmonics in the $n$ range from 101 to 1100. The reason why the rectangle window function was used is that a time-varying mean value is hardly subtracted from the output signal and it possibly has important components for reservoir computing. It is well recognized that this DFT procedure can lead to components at frequencies other than these harmonic frequencies and that quantitative analyses on sub-$f_\textup{IN}$ and interharmonics have not been yet established \cite{testa2007interharmonics}. Hence, the analysis with DFT spectra is phenomenological based on how the spectrum feature changes with the parameters $X$ and $\alpha$.  

\section{Physical Properties of Spin Dynamics at Detectors}
\begin{figure}[tb]
\includegraphics[width=\linewidth]{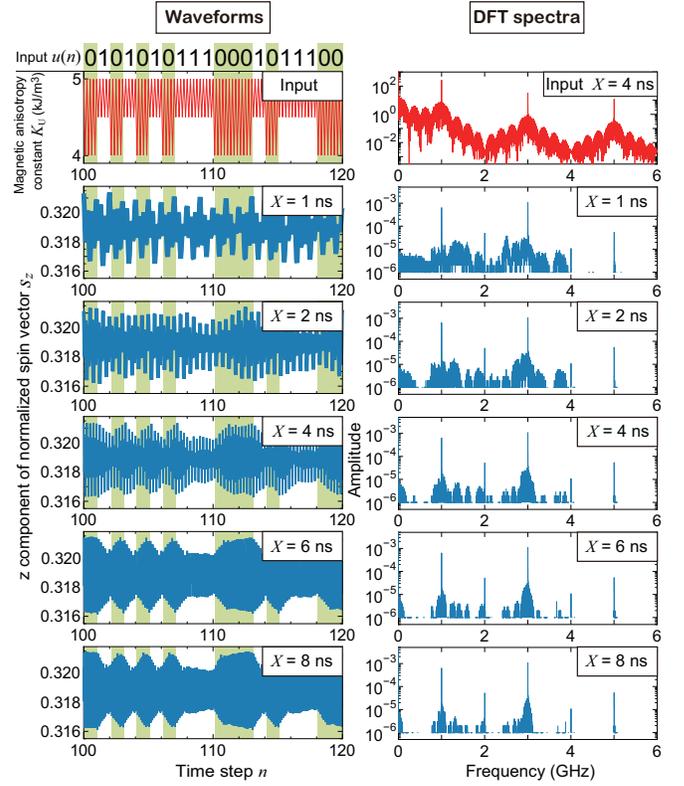}
\caption{\label{FIG_6}Red lines in the top two figures: a waveform and a DFT spectrum of input triangular pulses with two modulated amplitudes corresponding to bit sequence data, for which the transformation rule from the bit to amplitude of $K_\textup{U}$ is shown in Fig. \ref{FIG_5}(a). In the left-hand side, the unit of the horizontal axis is time step $n$, i.e., the time normalized by $X$ ns. In the right-hand side, the DFT spectrum calculated for $X$ = 4 ns is shown. Blue lines: Waveforms and DFT spectra of the resultant spin waves detected at Detector 1 when $\alpha$ = 0.1 and various time steps $X$ (= 1, 2, 4, 6, and 8 ns).  In the waveforms, the green ranges correspond to the bit value “0” ($\Delta K_\textup{u, 0}$), whereas the rest white ranges correspond to the bit value “1” ($\Delta K_\textup{u, 1}$). Note that the widths of the lines for $X$ = 1 and 2 ns are boroadened compared to those of the others to clearly see the outlines.}
\end{figure}

Figure \ref{FIG_6} shows waveforms and DFT spectra obtained with $\alpha$ = 0.1 and various $X$ (= 1, 2, 4, 6, and 8 ns), where the unit of the horizontal axis for the waveforms is time step $n$, red-line plots are results for the reservoir input signal with $X$ = 4 ns, and blue-line plots are results for reservoir output signals at Detector 1. In the waveforms, the green sections correspond to input value of 0 ($\Delta K_\textup{u, 0}$), whereas the rest white sections correspond to input value of 1($\Delta K_\textup{u, 1}$). In the DFT spectrum of the reservoir input signal in the upper-right panel, whereas the peaks at 1.0, 3.0, and 5.0 GHz are dominant, the background amplitude shows a gradual decrease with increasing frequency as well as a small oscillation with a constant period. The odd-number harmonics are consistent with the Fourier series expansion of sequential triangular pulses and the gradual decrease is simply interpreted as the random change in $\Delta K_\textup{u}$ with $n$. On the other hand, the period of the small oscillation is 0.25 GHz that corresponds to 1/$X$ = 1/4 $\textup{ns}^{-1}$. Since this relation in oscillation was also confirmed in the DFT spectra for input signals with other $X$ values (not shown here), it is attributable to the change in $\Delta K_\textup{u}$ with the step of $X$. 

In the reservoir output signals plotted with the blue lines, the following features are seen. The prominent feature is that nonlinear phenomena were obtained under all the conditions examined, since all the DFT spectra have peaked components at the even-number harmonic frequencies and components at the sub-$f_\textup{IN}$ and interharmonic frequencies. At $X$ = 1 ns, the amplitude of the waveform changes frequently with $n$ and its outline does not apparently respond to the reservoir input signal at the present time step $n$. As $X$ is increased, the outline of the waveform more clearly responds to the reservoir input signal at the present time step, in such a way that the amplitude of the waveform increases (decreases) after the bit transition from 1 to 0 (from 0 to 1). In the $n$ range of 107 – 114 (corresponds to an input sequence of [0, 1, 1, 1, 0, 0, 0]), the waveforms for $X$ = 2, 4, 6, and 8 ns show an almost complete change from the maximum to minimum amplitudes, and vice versa, with similar relaxation times. In the $n$ range of 100 – 107 (corresponds to an input sequence of [0, 1, 0, 1, 0, 1, 0]), the outlines of the waveforms slightly differ from each other. Considering the similar relaxation times in the $n$ range of 107 – 114, the difference in the outline of the wavefrom between $X$ = 2, 4, 6, and 8 ns mainly arises from the difference in the time step length $X$. On the other hand, the DFT spectrum changes with $X$. As $X$ is increased, the sub-$f_\textup{IN}$ and interharmonic components decrease, whereas the integer-number harmonic components remain almost unchanged. These features indicate that the most change in spin dynamics originates from nonlinear phenomena characterized by the sub-$f_\textup{IN}$ and interharmonic components.
\begin{figure}[tb]
\includegraphics[width=\linewidth]{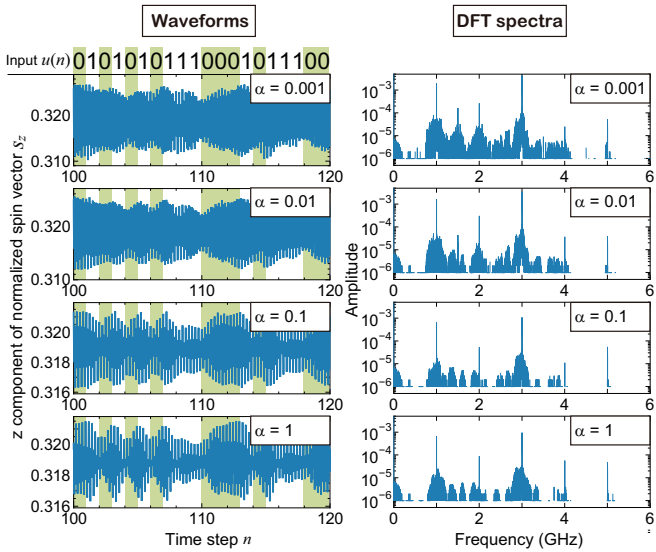}
\caption{\label{FIG_7}Waveforms and DFT spectra of reservoir output signals obtained at Detector 1 when various $\alpha$ values (= 0.001, 0.01, 0.1, and 1) and $X$ = 4 ns.}
\end{figure}

Figure \ref{FIG_7} shows waveforms and DFT spectra obtained with various $\alpha$ values (= 0.001, 0.01, 0.1, and 1) and $X$ = 4 ns, where blue lines are the results for reservoir output signals at Detector 1. In the $n$ range of 100 – 107 (corresponds to an input sequence of [0, 1, 0, 1, 0, 1, 0]), the outline of the waveform responds to the present input signal more clearly with increasing $\alpha$, while the amplitude becomes smaller. In the DFT spectra, when the sub-$f_\textup{IN}$ components and the relative ratios between the integer-number harmonic components are observed, they change little with $\alpha$. In contrast, the interhamonic components greatly increase with decreasing $\alpha$, which is similar to the feature with decreasing $X$ in Fig. \ref{FIG_6}. However, the spectrum for $\alpha$ = 0.001 and $X$ = 4 ns in Fig. \ref{FIG_7} differs from that for $\alpha$ = 0.1 and $X$ = 1 ns in Fig. \ref{FIG_6}. Thus, the effects of decreasing $\alpha$ and $X$ on the reservoir output signal are roughly similar, but they are not identical. This indicates that the change in the spin dynamics by $\alpha$ is different from that by $X$. 
\begin{figure}[tb]
\includegraphics[width=\linewidth]{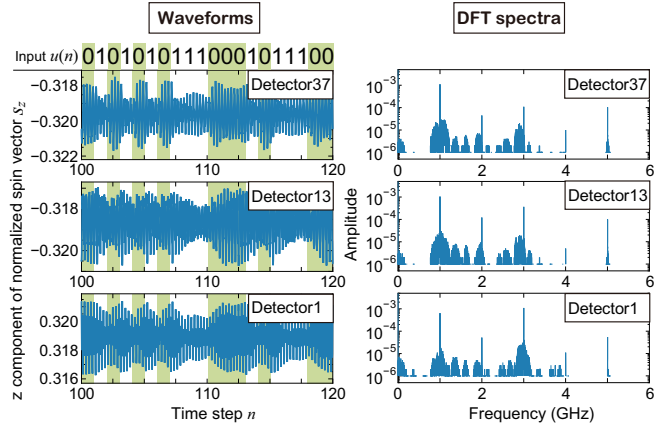}
\caption{\label{FIG_8}Waveforms and DFT spectra of reservoir output signals obtained at Detectors 37, 13, and 1 when $\alpha$ = 0.1 and $X$ = 4 ns.}
\end{figure}

Reservoir output signals at various detectors were observed to confirm the realization of various waveforms. Figure \ref{FIG_8} shows waveforms and DFT spectra at Detector 1, 13, and 37 obtained for $\alpha$ = 0.1 and $X$ = 4 ns. Whereas the three detectors are located at the nearest $x$ position from the input exciter, the $y$ positions are different, as shown in Fig. \ref{FIG_3}(c). The three waveforms are roughly similar, but not identical. The difference is also characterized by the DFT spectra, where each spectrum has a unique feature that is most simply confirmed in the relative ratios between the integer-number harmonic components. Thus, it was found that the three detectors have the different output waveforms even though they are located in the stripe domain structure without significant deviation, as seen in Fig. \ref{FIG_4}(a). In the same manner, when waveforms and DFT spectra at all the detectors were observed, they were unique for each detector.
\begin{figure}[tb]
\includegraphics[width=\linewidth]{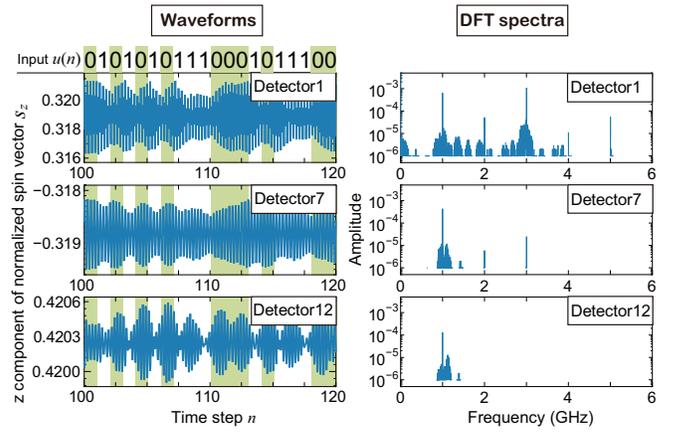}
\caption{\label{FIG_9}Waveforms and DFT spectra of reservoir output signals obtained at Detectors 1, 7, and 12 when $\alpha$ = 0.1 and $X$ = 4 ns.}
\end{figure}

Important phenomena were also found when waveforms at detectors in the same line along the $x$ axis were observed. Figure \ref{FIG_9} shows waveforms and DFT spectra at Detector 1, 7, and 12 when $\alpha$ = 0.1 and $X$ = 4 ns. With the increase of the detector number from 1 to 7 (the increase of the distance from the input exciter), the waveforms more ambiguously respond to the present input signal, which are confirmed by the valleys and peaks of the outline. Moreover, even when the waveform at Detector 1 is shifted along the horizontal axis and its amplitude is reduced, a waveform identical with that at Detector 7 cannot be created. Thus, it is very probable that the spin waves are nonlinearly changed while propagating due to nonlinear phenomena and reflection with multiple times. In the waveform at Detector 12, long-period beats with a width of $\sim$4 ns appear and the relation between the waveform and input signal is unclear. On the other hand, in the DFT spectra, as the detector number increases, the components at higher frequencies preferentially decrease and finally the components at around 1 GHz are present at Detector 12. Thus, the change in the waveform with increasing distance from the input exciter partially originates from the feature that spin waves propagating in long distances are inevitably determined by the stripe domain structure.

\section{Postprocess of Reservoir Output Signals}
\begin{figure}[tb]
\includegraphics[width=6.0 cm]{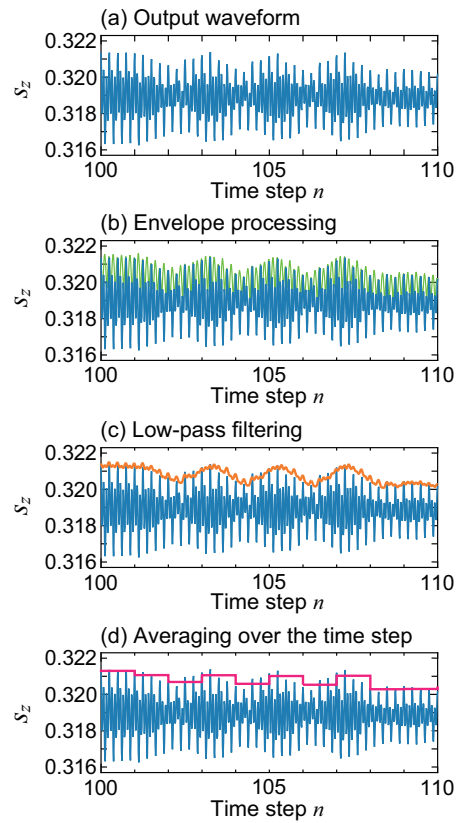}
\caption{\label{FIG_10} Post processing of a reservoir output signal, which is executed in the order of (a) reservoir output signal, (b) an envelope processing, (c) a low-pass filtering, and (d) averaging over the time step $X$. Blue lines in (a)$-$(d) are the same signal obtained at Detector 1 when $\alpha$ = 0.1 and $X$ = 4 ns. Colored lines in (b)$-$(d) are the signals after the processing. The $x_i(n)$ value in (d) is assigned to $y_i(n) (i = 1, \cdots, 72)$ in $\mathbf{y}(n)$.}
\end{figure}

In the readout part in Fig. \ref{FIG_1}, the reservoir output signals $x_i(t) (i = 1, \cdots, 72)$ are converted to a 72-dimensional output vector $\mathbf{y}(n)$ at each time step $n$. More specifically, $x_i(t)$ in a time step period are summarized into one value that reflects a characteristic response to the input signals from near past to $n$. As seen in the DFT spectra, the reservoir output signals have oscillations whose fundamental time period is smaller than $X$ and their envelopes are expected to reflect characteristic responses. Moreover, instantaneous changes should be excluded since they can be noise that does not reflect characteristic responses. Based on these considerations, the following procedure of the post signal processing was used, as shown in Figs. \ref{FIG_10}(a)$-$(d). First, an envelope processing and a subsequent low-pass filtering were performed for the reservoir output signals $x_i(t) (i = 1, \cdots, 72)$, as shown in Figs. \ref{FIG_10}(b) and (c). In the low-pass filtering, the cut-off frequency $f_\textup{c}$ defined by $-$10 dB is 1.3 GHz. Then, the resultant signal at each detector was averaged over the time period of $n$, i.e., $X$, as shown in Fig. \ref{FIG_10}(d). Finally, the averaged value at Detector $i$ was assigned to $y_i(n) (i = 1, \cdots, 72)$ in $\mathbf{y}(n)$. Clearly, the post signal processing extracts information in the low frequency range $\leq f_\textup{c}$. Nonetheless, spin waves in a wide frequency range are probably related to $\mathbf{y}(n)$ since they can dynamically interact with each other through nonlinear phenomena. It should be noted that the above postprocess procedure is feasible for implementation since each signal processing is widely used one. 

\section{Reservoir Computing}
\subsection{\label{XOR}Temporal XOR Problems}
\begin{figure}[tb]
\includegraphics[width=7.0 cm]{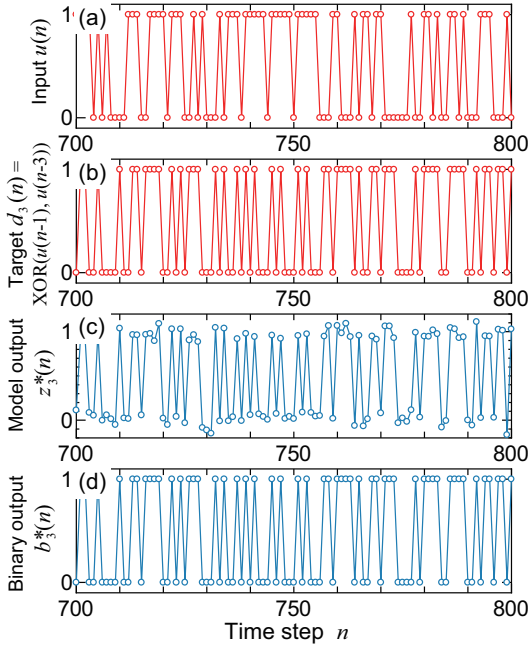}
\caption{\label{FIG_11}Singals in solving a XOR problem using 72-dimensional $\mathbf{y}(n)$ with the target output $d_3(n) = {\rm XOR}(u(n-1), u(n-3))$ (corresponding to delay $k$ = 3) when  $\alpha$ = 0.1 and $X$ = 4 ns. Red open circles are the input binary data and blue open circles represent the calculated values in the readout part. (a) Input $u(n)$ that has a bit value (0 or 1) changing randomly with $n$. (b) Target signal $d_3(n) = {\rm XOR}(u(n-1), u(n-3))$. (c) Model output $z_3^*(n)$. (d) Binary output $b_3^*(n)$ calculated from $z_3^*(n)$ with a threshold of 0.5. }
\end{figure}

\begin{figure}[tb]
\includegraphics[width=\linewidth]{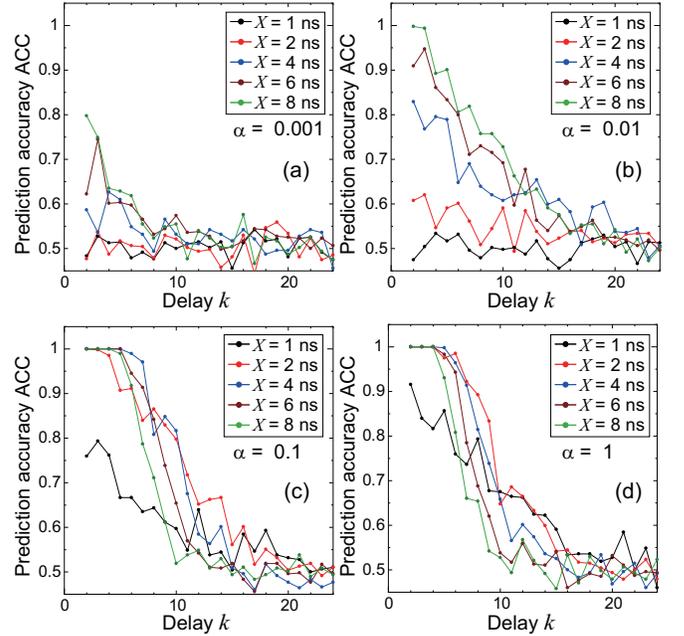}
\caption{\label{FIG_12}Prediction accuracy ACC plotted against delay $k$, which were estimated by solving XOR problems using 72-dimensional $\mathbf{y}(n)$ with $d_k(n) = {\rm XOR}(u(n-1), u(n-k))$ for delay $k = 2,\cdots,24$: $\alpha$ = (a) 0.001, (b) 0.01, (c) 0.1, and (d) 1, where black, red, blue, brown, and green dots represent the results for $X$ = 1, 2, 4, 6, and 8 ns, respectively, and colored lines connecting nearest-neighbor dots are guides.}
\end{figure}

Reservoir computing was performed using 72-dimensional $\mathbf{y}(n)$ for various conditions with $\alpha$ (= 0.001, 0.01, 0.1, and 1) and $X$ (= 1, 2, 4, 6, and 8 ns). Hereafter, $\mathbf{y}(n)$ has 72 dimensions unless otherwise noted.
To evaluate the device capability for nonlinear transformation of the input signal to $\mathbf{y}(n)$, temporal XOR problems were solved with the target output:
\begin{eqnarray}
 d_k(n) = {\rm XOR}(u(n-1), u(n-k)) \\ \hspace{2cm} \mbox{for delay} \, k &=& 2,\ldots, 24.\nonumber
\end{eqnarray} 
To solve the temporal XOR problem with delay $k$, the signal transformation by the reservoir requires nonlinearity and a memory of past $k$ inputs. We assume $k \ge 2$ to eliminate the trivial case of $k=1$ where $d_k(n) = 0$ for all $n$. Here, to evaluate the capability, the index CP is defined by the maximum value of $k$ such that 100$\%$ prediction accuracy (ACC = 1) is maintained when $k$ is increased. First, signals in the computation flow were observed by one example. Figures \ref{FIG_11}(a), (b), (c), and (d) show the input $u(n)$, the target output $d_3(n)$, the model output $z_3^*(n)$, and the binary output $b^*(n)$, respectively, where $\alpha$ = 0.1, $X$ = 4 ns, open circles are the values at $n$, and colored lines connecting the nearest neighbor open circles are guides. In this case, $d_3(n) = b_3^*(n)$ in the entire $n$ range, meaning ACC = 1. Figures \ref{FIG_12}(a), (b), (c), and (d) show ACC calculated for $\alpha$ = 0.001, 0.01, 0.1, and 1, respectively, plotted against $k$, where black, red, blue, brown, and green dots represent the results for $X$ = 1, 2, 4, 6, and 8 ns, respectively. When $\alpha$ = 0.001 in Fig. \ref{FIG_12}(a), all the curves do not have ACC = 1 and immidiately decrease to the minimum ACC = 0.5 with increasing $k$, in which the decreasing slope changes little with $X$. When $\alpha$ is increased from 0.001 to 0.01, ACC for each $X$ becomes larger at $k$ = 2 and the decreasing slope in each curve becomes more steep except $X$ = 1 ns, as shown in Fig. \ref{FIG_12}(b). However, since all the curves do not have ACC = 1, CP was not obtained. When $\alpha$ is increased from 0.01 to 0.1, the features of the curves are highly improved, as shown in Fig. \ref{FIG_12}(c). The curves for $X$ = 2, 4, 6, and 8 ns have ACC = 1 in $k$ $\geq$ 2, which are confirmed by the plateau in the low $k$ range. The curves for $X$ = 4, 6, and 8 ns have similar decreasing slopes. When $\alpha$ is increased from 0.1 to 1, the feature of the curve for $X$ = 2 ns is slightly improved, whereas the features of the curves for $X$ = 4, 6, and 8 ns are almost unchanged, as shown in Fig. \ref{FIG_12}(d). Table \ref{Table} lists CP for all the $\alpha$ and $X$ conditions, in which CP $\geq$ 4 was obtained for 7 conditions and the highest CP = 5 was obtained for $\alpha$ = 0.1 with $X$ = 4 and 6 ns.
\begin{figure}[tb]
\includegraphics[width=\linewidth]{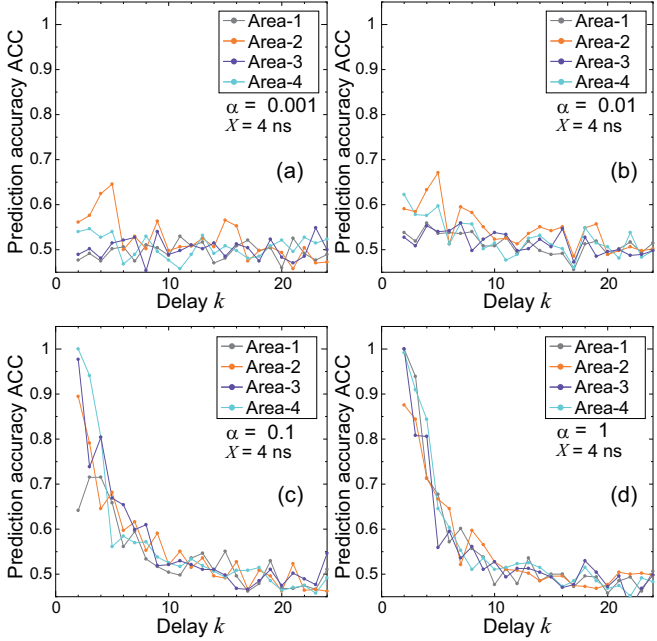}
\caption{\label{FIG_13}Prediction accuracy ACC plotted against delay $k$, which were estimated by solving XOR problems using 18-dimensional $\mathbf{y}(n)$ with $d_k(n) = {\rm XOR}(u(n-1), u(n-k))$ for delay $k = 2,\cdots,24$: $\alpha$ = (a) 0.001, (b) 0.01, (c) 0.1, and (d) 1, where grey, orange, pale-blue, and pale-green dots represent the results with $X$ = 4 ns for Area-1, -2, -3, and -4, respectively, and colored lines connecting nearest-neighbor dots are guides.}
\end{figure}

\begin{table*}[tb]
    \caption{\label{Table}Summaries of the computing capabilities $\textup{CP}$, $\textup{MC}^\textup{acc}$, and MC obtained for various $\alpha$ and $X$ values. CP for the temporal XOR ploblems is defined by the maximum value of delay $k$ such that 100$\%$ prediction accuracy (ACC = 1) is maintained when $k$ is increased, where bars in CP colums represent N/A. $\textup{MC}^\textup{acc}$ for the delay tasks is defined by the maximum value of delay $k$ such that 100$\%$ prediction accuracy (ACC = 1) is maintained when $k$ is increased. MC for the delay tasks is defined by Eq. (13).}
    \newcommand{\bhline}[1]{\noalign{\hrule height #1}}  
\newcommand{\bvline}[1]{\vrule width #1}
        \begin{center}
         \scalebox{1.0}{
            \begin{tabular}{|c||c|c|c||c|c|c||c|c|c||c|c|c|}\hline
            \multicolumn{1}{|c||}{} & \multicolumn{3}{c||}{$\alpha$ = 0.001}& \multicolumn{3}{c||}{$\alpha$ = 0.01}& \multicolumn{3}{c||}{$\alpha$ = 0.1}& \multicolumn{3}{c|}{$\alpha$ = 1}\rule[0mm]{0mm}{3mm}\\
            \cline{1-13}
            \multicolumn{1}{|c||}{} & \multicolumn{1}{c|}{$\textup{XOR problem}$} & \multicolumn{2}{c||}{$\textup{Delay task}$} & \multicolumn{1}{c|}{$\textup{XOR problem}$} & \multicolumn{2}{c||}{$\textup{Delay task}$} & \multicolumn{1}{c|}{$\textup{XOR problem}$} & \multicolumn{2}{c||}{$\textup{Delay task}$}&\multicolumn{1}{c|}{$\textup{XOR problem}$} & \multicolumn{2}{c|}{$\textup{Delay task}$} \rule[0mm]{0mm}{3.5mm}\\
            \cline{1-13}
            $X$ & $\textup{CP}$ & $\textup{MC}^\textup{acc}$ & MC & $\textup{CP}$ & $\textup{MC}^\textup{acc}$ & MC & $\textup{CP}$ & $\textup{MC}^\textup{acc}$ & MC & $\textup{CP}$ & $\textup{MC}^\textup{acc}$ & MC  \rule[0mm]{0mm}{4mm}\\ \bhline{1.0pt}
            1 ns & $-$ & 0 &14 & $-$ & 0 & 36 & $-$ & 15 & 31 & $-$ & 18 & 28 \rule[0mm]{0mm}{3mm}\\ \hline
            2 ns & $-$ & 0 & 19 & $-$ &   9 & 42 & 2 & 16 & 24 & 4 &  13 & 21  \rule[0mm]{0mm}{3mm}\\ \hline
            4 ns & $-$ & 1 & 20 & $-$ & 15 & 32 & 5 & 12 & 16 & 4 & 11 & 15  \rule[0mm]{0mm}{3mm}\\ \hline
            6 ns & $-$ & 3 & 18 & $-$ & 15 & 26 & 5 & 10 & 13 & 4 &  10 & 13 \rule[0mm]{0mm}{3mm}\\ \hline
            8 ns & $-$ & 5 & 18 & $-$ & 13 & 23 & 4 & 9 & 10 & 4 &  8 & 11  \rule[0mm]{0mm}{3mm}\\ \hline
        \end{tabular}
        }
    \end{center}
\end{table*}

So far, the reservoir computing was performed using 72-dimensional $\mathbf{y}(n)$ from 72 detectors. Here, the detectors are divided into 4 groups named Area-1, -2, -3, and -4, as shown in Fig. \ref{FIG_3}(c), and then the temporal XOR problems were solved using 18 detectors in each area, when $\alpha$ was varied and $X$ = 4 ns. In consequence, $\mathbf{y}(n)$ for each computation is a 18-dimensional vector. Figures \ref{FIG_13}(a), (b), (c), and (d) show ACC calculated for $\alpha$ = 0.001, 0.01, 0.1, and 1, respectively, plotted against $k$, where grey, orange, pale-blue, and pale-green dots represent the results for Area-1, -2, -3, and -4, respectively. At each $\alpha$ value, the curves for Area-1, -2, -3, and -4 are similar to each other and have the smaller values than the curve for $X$ = 4 ns in Figs. \ref{FIG_12}(a), (b), (c), and (d). The results clearly indicate that the $\mathbf{y}(n)$ dimension effective for reservoir computing increases with increasing number of the detector, i.e., higher dimensionality was efficiently obtained by the increase of the reservoir output signals with various waveforms. This finding also leads to a conclusion that the periodic arrangement of detectors in Fig. \ref{FIG_3}(c) works well for the realization of high dimensionality in $\mathbf{y}(n)$, which is advantageous for feasible implementation of the device.

Here, we discuss the effectiveness of the damping regions at the left and right boundaries. If spin waves are significantly reflected from these boundaries, they also contribute to the output signals at the detectors, particularly, in Area-1 and -4. This is because spin waves propagating from the input exciter can interfere with reflected spin waves that have memories in further past. In this situation, ACC values for Area-1 and -4 can differ from those for Area-2 and -3. On the other hand, the curves in Figs. \ref{FIG_13}(a), (b), (c), and (d) do not have a clear Area-dependence at each $\alpha$ value. Thus, the results indicate that spin waves reflected from the left and right boundaries were satisfactorily suppressed.

\subsection{Delay Tasks and Memory Capacity}
Memory capacity was estimated by solving delay tasks with the target output:
\begin{eqnarray}
d_k(n) = u(n-k) \hspace{0.5cm} \mbox{for delay} \, k = 1,\ldots, 72.
\end{eqnarray}
In this study, we evaluated memory capacity by two standards. The first one is MC that is defined by the widely-used formula as follows \cite{jaeger2002short}:

\begin{eqnarray}
\textup{MC} &=& \sum_{k = 1}^{72}{\gamma}^2(k) \nonumber \\
&=& \sum_{k = 1}^{72}\frac{\textup{Cov}^2(u(n-k), b_k^*(n))}{\textup{Var}(u(n))\textup{Var}(b_k^*(n))}, 
\end{eqnarray}
where $\gamma$, Cov, and Var denote the coefficient of determination, covariance, and variance, respectively, and $b_k^*(n)$ is the binary output at $n$ in the delay task with $k$, respectively. The main purpose of using this index is to study a fundamental feature of the reservoir device by comparing estimated values with those in other works. On the other hand, the second one is $\textup{MC}^\textup{acc}$ that is defined by the maximum value of $k$ such that 100$\%$ prediction accuracy (ACC = 1) is maintained when $k$ is increased. The purpose of using this index is to clarify the conditions for applications with high-accuracy and long-step memory, such as forecasting nonlinear autoregressive moving average (NARMA) time series \cite{atiya2000new}. It should be noted that $\textup{MC}^\textup{acc}$ has a value that is equal to or less than MC.
\begin{figure}[tb]
\includegraphics[width=\linewidth]{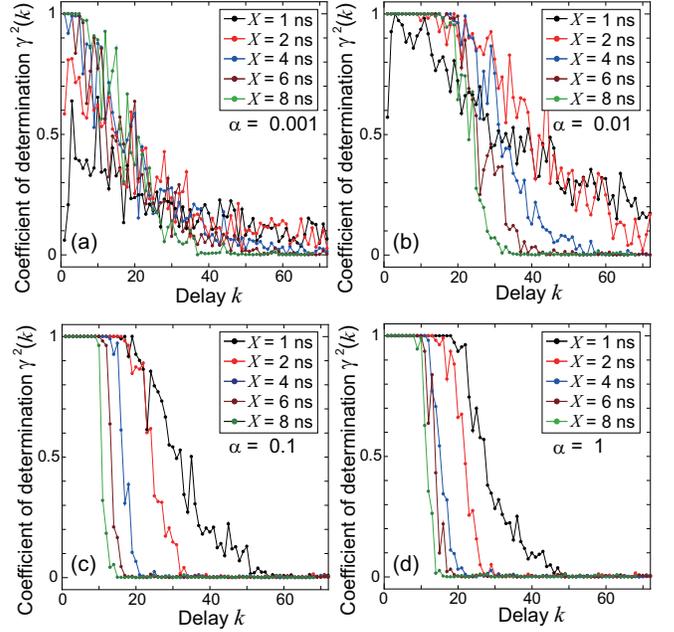}
\caption{\label{FIG_14} Coefficient of determination ${\gamma}^2(k)$ plotted against delay $k$, which were estimated by solving delay tasks using 72-dimensional $\mathbf{y}(n)$ with the target output $d(n) = u(n-k)$ for delay $k=1,\cdots,72$: $\alpha$ = (a) 0.001, (b) 0.01, (c) 0.1, and (d) 1, where black, red, blue, brown, and green dots represent the results for $X$ = 1, 2, 4, 6, and 8 ns, respectively, and colored lines connecting nearest-neighbor dots are guides.}
\end{figure}

Figures \ref{FIG_14}(a), (b), (c), and (d) show ${\gamma}^2(k)$ calculated for $\alpha$ = 0.001, 0.01, 0.1, and 1, respectively, plotted against delay $k$, where black, red, blue, brown, and green dots represent the results for $X$ = 1, 2, 4, 6, and 8 ns, respectively. When $\alpha$ = 0.001 in Fig. \ref{FIG_14}(a), the curves for all the $X$ values monotonically decrease with increasing $k$ and their shapes are similar to each other. The estimated values of $\textup{MC}^\textup{acc}$ for $X$ = 1, 2, 4, 6, and 8 ns are 0, 0, 1, 3, and 5, respectively. When $\alpha$ = 0.01 in Fig. \ref{FIG_14}(b), the curves show different features depending on $X$. As $k$ is increased from 0, the curves except $X$ = 1 ns show ${\gamma}^2(k)$ = 1 at first and then they monotonically decrease. It was found that  $\textup{MC}^\textup{acc}$ are not related to the fact that the decreasing slope becomes more steep with increasing $X$: $\textup{MC}^\textup{acc}$ for $X$ = 1, 2, 4, 6, and 8 ns are 0, 9, 15, 15, and 13, respectively. When $\alpha$ = 0.1 in Fig. \ref{FIG_14}(c), all the curves are clearly different. As $k$ is increased from 0, all the curves show ${\gamma}^2(k)$ = 1 at first, whereas the $k$ position where the curve bends toward 0 becomes lower with increasing $X$. Owing to this feature, $\textup{MC}^\textup{acc}$ approximately decreases with increasing $X$: The estimated values of $\textup{MC}^\textup{acc}$ for $X$ = 1, 2, 4, 6, and 8 ns are 15, 16, 12, 10, and 9, respectively. When $\alpha$ = 1 in Fig. \ref{FIG_14}(d), the overall feature is similar to that in Fig. \ref{FIG_14}(c), whereas $\textup{MC}^\textup{acc}$ for $X$ = 1 ns is slightly improved: The estimated values of $\textup{MC}^\textup{acc}$ for $X$ = 1, 2, 4, 6, and 8 ns are 18, 13, 11, 10, and 8, respectively.

The results for $\textup{MC}^\textup{acc}$ and MC are summarized in Table \ref{Table}. When $\alpha$ = 0.001, MC is considerably larger than $ \textup{MC}^\textup{acc}$ in all the $X$ cases, which arise from the gentle decreases in ${\gamma}^2(k)$ with increasing $k$ in Fig. \ref{FIG_14}(a). When $\alpha$ = 0.01, MC is also considerably larger than $\textup{MC}^\textup{acc}$ in the cases of $X$ = 1 and 2 ns. This is because the decreasing slopes for $X$ = 1 and 2 ns are more gradual than those for other $X$ values, as seen in Fig. \ref{FIG_14}(b). Thus, under the $\alpha$ condition that the decreasing slope depends significantly on $X$, larger $\textup{MC}^\textup{acc}$ does not always lead to larger MC, i.e., $\textup{MC}^\textup{acc}$ is not correlated with MC. For $\alpha$ = 0.1 and 1, where the decreasing slopes for all the $X$ values are similar to each other, the estimated values of $\textup{MC}^\textup{acc}$ are reasonably correlated with MC and, in a wide range of $X$, $\textup{MC}^\textup{acc}$ $>$ 10 and MC $>$ 13 were achieved. These MC values are comparable to those obtained in the reservoir system with a Mackey-Glass nonlinearity and a feedback loop \cite{appeltant2012reservoir} and, to the best of our knowledge, the maximum MC value of 21 is the highest ever in the on-chip electronic systems \cite{yamaguchi2020periodic, furuta2018macromagnetic, kanao2019reservoir, nako2020proposal}.

\section{Discussion}
Toward deep understanding of physical reservoir computing, it is worthwhile to discuss the relationship between the reservoir-computing capabilities CP and $\textup{MC}^\textup{acc}$, with suggestions on the physical properties of the spin waves behind the reservoir computing. The preceding sections have mainly studied on how the physical properties of the spin waves and the reservoir-computing capabilities are changed by the two parameters: the device parameter $\alpha$ and the input signal parameter $X$. 

In the evaluation of the capability for the temporal XOR problems, the $\alpha$ and $X$ conditions are narrow for CP $\geq$ 2: $X$ = 2, 4, 6, and 8 ns with $\alpha$ = 0.1 and $X$ = 2, 4, 6, and 8 ns with $\alpha$ = 1. Thus, $\alpha$ is the primal parameter for solving the temporal XOR problems. Since the decreases of $\alpha$ and $X$ result in the enhancement of nonlinear phenomena characterized by the sub-$f_\textup{IN}$ and interhamonic components, excessive nonlinear phenomena degrade CP. It may be considered that richer nonlinear phenomena are preferable for linearly-inseparable problems as the temporal XOR problems, but the results indicate that there are effective and obstructive nonlinear phenomena for reservoir computing. In the evaluation of the memory capacities, the $\textup{MC}^\textup{acc}$ values higher than 10 were obtained under the following conditions: $X$ = 4, 6, and 8 ns with $\alpha$ = 0.01, $X$ = 1, 2, 4, and 6 ns with $\alpha$ = 0.1, and $X$ = 1, 2, 4, and 6 ns with $\alpha$ = 1. Thus, the $X$ and $\alpha$ conditions are broad even though the DFT spectra are clearly changed with the $X$ and $\alpha$ values in those ranges. Considering CP and $\textup{MC}^\textup{acc}$, the device and input conditions for high-accuracy computing is more restrictive for the temporal XOR problems.

On the other hand, as pointed out previously, whereas the post signal processing in the readout part extracts the characteristics only in the frequency range lower than $f_\textup{c}$ = 1.3 GHz, the computation capabilities strongly depend on $\alpha$ and $X$. This is clear evidence that the characteristics in the high and low frequency ranges dynamically interact with each other through nonlinear phenomena since the influence of the change in $\alpha$ and $X$ was mostly observed in the frequency range higher than $f_\textup{c}$. Since nonlinear phenomena were characterized only by the DFT spectra that do not directly capture dynamic behavior, further analyses would be necessary. Nonetheless, the above finding is noteworthy for understanding of physical reservoir computing.

When the discussion is focused on the cases of $\alpha$ = 0.1 and 1, $\textup{MC}^\textup{acc}$ almost monotonically decreases and CP appears with increasing $X$ at each $\alpha$, as confirmed in Table \ref{Table}. This may indicate a tradeoff between $\textup{MC}^\textup{acc}$ and CP, as pointed out for the echo state network \cite{inubushi2017reservoir}. In the present study, since the CP values are not so high, the change in CP with $X$ is not significant. This probably arises from the fact that the $\mathbf{y}(n)$ dimension is 72. Considering the result that the CP value becomes higher with increasing number of the output electrodes in Figs. \ref{FIG_12} and \ref{FIG_13}, it is expected that much higher CP values can be obtained with higher-dimensional $\mathbf{y}(n)$ that is generated by a reservoir device with larger numbers of output electrodes. This achievement may lead to a significant change in CP with $X$, which can provide further insight into the relation between $\textup{MC}^\textup{acc}$ and CP.

\section{Summary}
In the first half of the numerical experiments, it was shown how the physical properties of the spin waves were changed by the two parameters: the device parameter $\alpha$ and the input signal parameter $X$. From the characterization by the DFT spectra, it was found that nonlinear phenomena were obtained in all the $\alpha$ and $X$ conditions examined here, which is mainly attributable to the stripe domain structure and unstable spin waves under the low external magnetic field. As $\alpha$ and $X$ are decreased, nonlinear phenomena more frequently occur, which was mostly characterized by the increase of the sub-$f_\textup{IN}$ and interhamonic components in the DFT spectra. In the latter half of the numerical experiments, the reservoir computing was performed to evaluate CP by solving the temporal XOR problems as well as $\textup{MC}^\textup{acc}$ and MC by solving the delay tasks. In the temporal XOR problems, CP $\geq$ 4 was obtained under the following conditions: $X$ = 4, 6, and 8 ns with $\alpha$ = 0.1 and $X$ = 2, 4, 6, and 8 ns with $\alpha$ = 1. Since the sub-$f_\textup{IN}$ and interhamonic components for $\alpha$ = 0.1 and 1 are smaller than those for $\alpha$ = 0.001 and 0.01, moderate suppression of nonlinear physical phenomena was found to be a key to such high CP values. Among the conditions for CP $\geq$ 4, $\textup{MC}^\textup{acc}$ and MC are relatively high and decrease with increasing $X$. The maximum value of $\textup{MC}^\textup{acc}$ is 13 at $\alpha$ = 1 and $X$ = 2 ns condition and the value of MC is 21 at the same condition.  Therefore, this study demonstrated that the spin-wave-based reservoir computing system simultaneously achieves the relatively-high CP, $\textup{MC}^\textup{acc}$, and MC values among electronic reservoir-computing hardware systems.

Since the spin-wave-based reservoir device utilizes the signal transmission through a continuous magnetic garnet film and the signal excitation/detection by input/output electrodes on the top, it is expected that the computing capabilities of the reservoir system can be further improved and various functionalities for edge-computing applications can be realized just by changing the number, selection, and the arrangement of input/output electrodes, an external magnetic field, and the magnetic domain structure. Such scalability also can offer a rich opportunity for further understanding of the underlying nonlinear physics.
\begin{acknowledgments}
This work is based on results obtained from a project,
JPNP16007, commissioned by the New Energy and Industrial Technology Development Organization (NEDO).
\end{acknowledgments}


%\nocite{*}


\begin{thebibliography}{42}%
\makeatletter
\providecommand \@ifxundefined [1]{%
 \@ifx{#1\undefined}
}%
\providecommand \@ifnum [1]{%
 \ifnum #1\expandafter \@firstoftwo
 \else \expandafter \@secondoftwo
 \fi
}%
\providecommand \@ifx [1]{%
 \ifx #1\expandafter \@firstoftwo
 \else \expandafter \@secondoftwo
 \fi
}%
\providecommand \natexlab [1]{#1}%
\providecommand \enquote  [1]{``#1''}%
\providecommand \bibnamefont  [1]{#1}%
\providecommand \bibfnamefont [1]{#1}%
\providecommand \citenamefont [1]{#1}%
\providecommand \href@noop [0]{\@secondoftwo}%
\providecommand \href [0]{\begingroup \@sanitize@url \@href}%
\providecommand \@href[1]{\@@startlink{#1}\@@href}%
\providecommand \@@href[1]{\endgroup#1\@@endlink}%
\providecommand \@sanitize@url [0]{\catcode `\\12\catcode `\$12\catcode
  `\&12\catcode `\#12\catcode `\^12\catcode `\_12\catcode `\%12\relax}%
\providecommand \@@startlink[1]{}%
\providecommand \@@endlink[0]{}%
\providecommand \url  [0]{\begingroup\@sanitize@url \@url }%
\providecommand \@url [1]{\endgroup\@href {#1}{\urlprefix }}%
\providecommand \urlprefix  [0]{URL }%
\providecommand \Eprint [0]{\href }%
\providecommand \doibase [0]{https://doi.org/}%
\providecommand \selectlanguage [0]{\@gobble}%
\providecommand \bibinfo  [0]{\@secondoftwo}%
\providecommand \bibfield  [0]{\@secondoftwo}%
\providecommand \translation [1]{[#1]}%
\providecommand \BibitemOpen [0]{}%
\providecommand \bibitemStop [0]{}%
\providecommand \bibitemNoStop [0]{.\EOS\space}%
\providecommand \EOS [0]{\spacefactor3000\relax}%
\providecommand \BibitemShut  [1]{\csname bibitem#1\endcsname}%
\let\auto@bib@innerbib\@empty
%</preamble>
\bibitem [{\citenamefont {Jaeger}(2001)}]{jaeger2001echo}%
  \BibitemOpen
  \bibfield  {author} {\bibinfo {author} {\bibfnamefont {H.}~\bibnamefont
  {Jaeger}},\ }\bibfield  {title} {\bibinfo {title} {The “echo state”
  approach to analysing and training recurrent neural networks-with an erratum
  note},\ }\href@noop {} {\bibfield  {journal} {\bibinfo  {journal} {Bonn,
  Germany: German National Research Center for Information Technology GMD
  Technical Report}\ }\textbf {\bibinfo {volume} {148}},\ \bibinfo {pages} {13}
  (\bibinfo {year} {2001})}\BibitemShut {NoStop}%
\bibitem [{\citenamefont {Maass}\ \emph {et~al.}(2002)\citenamefont {Maass},
  \citenamefont {Natschl{\"a}ger},\ and\ \citenamefont
  {Markram}}]{maass2002real}%
  \BibitemOpen
  \bibfield  {author} {\bibinfo {author} {\bibfnamefont {W.}~\bibnamefont
  {Maass}}, \bibinfo {author} {\bibfnamefont {T.}~\bibnamefont
  {Natschl{\"a}ger}},\ and\ \bibinfo {author} {\bibfnamefont {H.}~\bibnamefont
  {Markram}},\ }\bibfield  {title} {\bibinfo {title} {Real-time computing
  without stable states: A new framework for neural computation based on
  perturbations},\ }\href@noop {} {\bibfield  {journal} {\bibinfo  {journal}
  {Neural computation}\ }\textbf {\bibinfo {volume} {14}},\ \bibinfo {pages}
  {2531} (\bibinfo {year} {2002})}\BibitemShut {NoStop}%
\bibitem [{\citenamefont {Shi}\ \emph {et~al.}(2016)\citenamefont {Shi},
  \citenamefont {Cao}, \citenamefont {Zhang}, \citenamefont {Li},\ and\
  \citenamefont {Xu}}]{shi2016edge}%
  \BibitemOpen
  \bibfield  {author} {\bibinfo {author} {\bibfnamefont {W.}~\bibnamefont
  {Shi}}, \bibinfo {author} {\bibfnamefont {J.}~\bibnamefont {Cao}}, \bibinfo
  {author} {\bibfnamefont {Q.}~\bibnamefont {Zhang}}, \bibinfo {author}
  {\bibfnamefont {Y.}~\bibnamefont {Li}},\ and\ \bibinfo {author}
  {\bibfnamefont {L.}~\bibnamefont {Xu}},\ }\bibfield  {title} {\bibinfo
  {title} {Edge computing: Vision and challenges},\ }\href@noop {} {\bibfield
  {journal} {\bibinfo  {journal} {IEEE internet of things journal}\ }\textbf
  {\bibinfo {volume} {3}},\ \bibinfo {pages} {637} (\bibinfo {year}
  {2016})}\BibitemShut {NoStop}%
\bibitem [{\citenamefont {Abbas}\ \emph {et~al.}(2017)\citenamefont {Abbas},
  \citenamefont {Zhang}, \citenamefont {Taherkordi},\ and\ \citenamefont
  {Skeie}}]{abbas2017mobile}%
  \BibitemOpen
  \bibfield  {author} {\bibinfo {author} {\bibfnamefont {N.}~\bibnamefont
  {Abbas}}, \bibinfo {author} {\bibfnamefont {Y.}~\bibnamefont {Zhang}},
  \bibinfo {author} {\bibfnamefont {A.}~\bibnamefont {Taherkordi}},\ and\
  \bibinfo {author} {\bibfnamefont {T.}~\bibnamefont {Skeie}},\ }\bibfield
  {title} {\bibinfo {title} {Mobile edge computing: A survey},\ }\href@noop {}
  {\bibfield  {journal} {\bibinfo  {journal} {IEEE Internet of Things Journal}\
  }\textbf {\bibinfo {volume} {5}},\ \bibinfo {pages} {450} (\bibinfo {year}
  {2017})}\BibitemShut {NoStop}%
\bibitem [{\citenamefont {Tanaka}\ \emph {et~al.}(2019)\citenamefont {Tanaka},
  \citenamefont {Yamane}, \citenamefont {H{\'e}roux}, \citenamefont {Nakane},
  \citenamefont {Kanazawa}, \citenamefont {Takeda}, \citenamefont {Numata},
  \citenamefont {Nakano},\ and\ \citenamefont {Hirose}}]{tanaka2019recent}%
  \BibitemOpen
  \bibfield  {author} {\bibinfo {author} {\bibfnamefont {G.}~\bibnamefont
  {Tanaka}}, \bibinfo {author} {\bibfnamefont {T.}~\bibnamefont {Yamane}},
  \bibinfo {author} {\bibfnamefont {J.~B.}\ \bibnamefont {H{\'e}roux}},
  \bibinfo {author} {\bibfnamefont {R.}~\bibnamefont {Nakane}}, \bibinfo
  {author} {\bibfnamefont {N.}~\bibnamefont {Kanazawa}}, \bibinfo {author}
  {\bibfnamefont {S.}~\bibnamefont {Takeda}}, \bibinfo {author} {\bibfnamefont
  {H.}~\bibnamefont {Numata}}, \bibinfo {author} {\bibfnamefont
  {D.}~\bibnamefont {Nakano}},\ and\ \bibinfo {author} {\bibfnamefont
  {A.}~\bibnamefont {Hirose}},\ }\bibfield  {title} {\bibinfo {title} {Recent
  advances in physical reservoir computing: A review},\ }\href@noop {}
  {\bibfield  {journal} {\bibinfo  {journal} {Neural Networks}\ }\textbf
  {\bibinfo {volume} {115}},\ \bibinfo {pages} {100} (\bibinfo {year}
  {2019})}\BibitemShut {NoStop}%
\bibitem [{\citenamefont {Larger}\ \emph {et~al.}(2012)\citenamefont {Larger},
  \citenamefont {Soriano}, \citenamefont {Brunner}, \citenamefont {Appeltant},
  \citenamefont {Guti{\'e}rrez}, \citenamefont {Pesquera}, \citenamefont
  {Mirasso},\ and\ \citenamefont {Fischer}}]{larger2012photonic}%
  \BibitemOpen
  \bibfield  {author} {\bibinfo {author} {\bibfnamefont {L.}~\bibnamefont
  {Larger}}, \bibinfo {author} {\bibfnamefont {M.~C.}\ \bibnamefont {Soriano}},
  \bibinfo {author} {\bibfnamefont {D.}~\bibnamefont {Brunner}}, \bibinfo
  {author} {\bibfnamefont {L.}~\bibnamefont {Appeltant}}, \bibinfo {author}
  {\bibfnamefont {J.~M.}\ \bibnamefont {Guti{\'e}rrez}}, \bibinfo {author}
  {\bibfnamefont {L.}~\bibnamefont {Pesquera}}, \bibinfo {author}
  {\bibfnamefont {C.~R.}\ \bibnamefont {Mirasso}},\ and\ \bibinfo {author}
  {\bibfnamefont {I.}~\bibnamefont {Fischer}},\ }\bibfield  {title} {\bibinfo
  {title} {Photonic information processing beyond turing: an optoelectronic
  implementation of reservoir computing},\ }\href@noop {} {\bibfield  {journal}
  {\bibinfo  {journal} {Optics express}\ }\textbf {\bibinfo {volume} {20}},\
  \bibinfo {pages} {3241} (\bibinfo {year} {2012})}\BibitemShut {NoStop}%
\bibitem [{\citenamefont {Paquot}\ \emph {et~al.}(2012)\citenamefont {Paquot},
  \citenamefont {Duport}, \citenamefont {Smerieri}, \citenamefont {Dambre},
  \citenamefont {Schrauwen}, \citenamefont {Haelterman},\ and\ \citenamefont
  {Massar}}]{paquot2012optoelectronic}%
  \BibitemOpen
  \bibfield  {author} {\bibinfo {author} {\bibfnamefont {Y.}~\bibnamefont
  {Paquot}}, \bibinfo {author} {\bibfnamefont {F.}~\bibnamefont {Duport}},
  \bibinfo {author} {\bibfnamefont {A.}~\bibnamefont {Smerieri}}, \bibinfo
  {author} {\bibfnamefont {J.}~\bibnamefont {Dambre}}, \bibinfo {author}
  {\bibfnamefont {B.}~\bibnamefont {Schrauwen}}, \bibinfo {author}
  {\bibfnamefont {M.}~\bibnamefont {Haelterman}},\ and\ \bibinfo {author}
  {\bibfnamefont {S.}~\bibnamefont {Massar}},\ }\bibfield  {title} {\bibinfo
  {title} {Optoelectronic reservoir computing},\ }\href@noop {} {\bibfield
  {journal} {\bibinfo  {journal} {Scientific reports}\ }\textbf {\bibinfo
  {volume} {2}},\ \bibinfo {pages} {287} (\bibinfo {year} {2012})}\BibitemShut
  {NoStop}%
\bibitem [{\citenamefont {Duport}\ \emph {et~al.}(2012)\citenamefont {Duport},
  \citenamefont {Schneider}, \citenamefont {Smerieri}, \citenamefont
  {Haelterman},\ and\ \citenamefont {Massar}}]{duport2012all}%
  \BibitemOpen
  \bibfield  {author} {\bibinfo {author} {\bibfnamefont {F.}~\bibnamefont
  {Duport}}, \bibinfo {author} {\bibfnamefont {B.}~\bibnamefont {Schneider}},
  \bibinfo {author} {\bibfnamefont {A.}~\bibnamefont {Smerieri}}, \bibinfo
  {author} {\bibfnamefont {M.}~\bibnamefont {Haelterman}},\ and\ \bibinfo
  {author} {\bibfnamefont {S.}~\bibnamefont {Massar}},\ }\bibfield  {title}
  {\bibinfo {title} {All-optical reservoir computing},\ }\href@noop {}
  {\bibfield  {journal} {\bibinfo  {journal} {Optics express}\ }\textbf
  {\bibinfo {volume} {20}},\ \bibinfo {pages} {22783} (\bibinfo {year}
  {2012})}\BibitemShut {NoStop}%
\bibitem [{\citenamefont {Brunner}\ \emph {et~al.}(2013)\citenamefont
  {Brunner}, \citenamefont {Soriano}, \citenamefont {Mirasso},\ and\
  \citenamefont {Fischer}}]{brunner2013parallel}%
  \BibitemOpen
  \bibfield  {author} {\bibinfo {author} {\bibfnamefont {D.}~\bibnamefont
  {Brunner}}, \bibinfo {author} {\bibfnamefont {M.~C.}\ \bibnamefont
  {Soriano}}, \bibinfo {author} {\bibfnamefont {C.~R.}\ \bibnamefont
  {Mirasso}},\ and\ \bibinfo {author} {\bibfnamefont {I.}~\bibnamefont
  {Fischer}},\ }\bibfield  {title} {\bibinfo {title} {Parallel photonic
  information processing at gigabyte per second data rates using transient
  states},\ }\href@noop {} {\bibfield  {journal} {\bibinfo  {journal} {Nature
  communications}\ }\textbf {\bibinfo {volume} {4}},\ \bibinfo {pages} {1}
  (\bibinfo {year} {2013})}\BibitemShut {NoStop}%
\bibitem [{\citenamefont {Appeltant}\ \emph {et~al.}(2011)\citenamefont
  {Appeltant}, \citenamefont {Soriano}, \citenamefont {Van~der Sande},
  \citenamefont {Danckaert}, \citenamefont {Massar}, \citenamefont {Dambre},
  \citenamefont {Schrauwen}, \citenamefont {Mirasso},\ and\ \citenamefont
  {Fischer}}]{appeltant2011information}%
  \BibitemOpen
  \bibfield  {author} {\bibinfo {author} {\bibfnamefont {L.}~\bibnamefont
  {Appeltant}}, \bibinfo {author} {\bibfnamefont {M.~C.}\ \bibnamefont
  {Soriano}}, \bibinfo {author} {\bibfnamefont {G.}~\bibnamefont {Van~der
  Sande}}, \bibinfo {author} {\bibfnamefont {J.}~\bibnamefont {Danckaert}},
  \bibinfo {author} {\bibfnamefont {S.}~\bibnamefont {Massar}}, \bibinfo
  {author} {\bibfnamefont {J.}~\bibnamefont {Dambre}}, \bibinfo {author}
  {\bibfnamefont {B.}~\bibnamefont {Schrauwen}}, \bibinfo {author}
  {\bibfnamefont {C.~R.}\ \bibnamefont {Mirasso}},\ and\ \bibinfo {author}
  {\bibfnamefont {I.}~\bibnamefont {Fischer}},\ }\bibfield  {title} {\bibinfo
  {title} {Information processing using a single dynamical node as complex
  system},\ }\href@noop {} {\bibfield  {journal} {\bibinfo  {journal} {Nature
  communications}\ } (\bibinfo {year} {2011})}\BibitemShut {NoStop}%
\bibitem [{\citenamefont {Du}\ \emph {et~al.}(2017)\citenamefont {Du},
  \citenamefont {Cai}, \citenamefont {Zidan}, \citenamefont {Ma}, \citenamefont
  {Lee},\ and\ \citenamefont {Lu}}]{du2017reservoir}%
  \BibitemOpen
  \bibfield  {author} {\bibinfo {author} {\bibfnamefont {C.}~\bibnamefont
  {Du}}, \bibinfo {author} {\bibfnamefont {F.}~\bibnamefont {Cai}}, \bibinfo
  {author} {\bibfnamefont {M.~A.}\ \bibnamefont {Zidan}}, \bibinfo {author}
  {\bibfnamefont {W.}~\bibnamefont {Ma}}, \bibinfo {author} {\bibfnamefont
  {S.~H.}\ \bibnamefont {Lee}},\ and\ \bibinfo {author} {\bibfnamefont {W.~D.}\
  \bibnamefont {Lu}},\ }\bibfield  {title} {\bibinfo {title} {Reservoir
  computing using dynamic memristors for temporal information processing},\
  }\href@noop {} {\bibfield  {journal} {\bibinfo  {journal} {Nature
  communications}\ }\textbf {\bibinfo {volume} {8}},\ \bibinfo {pages} {2204}
  (\bibinfo {year} {2017})}\BibitemShut {NoStop}%
\bibitem [{\citenamefont {Moon}\ \emph {et~al.}(2019)\citenamefont {Moon},
  \citenamefont {Ma}, \citenamefont {Shin}, \citenamefont {Cai}, \citenamefont
  {Du}, \citenamefont {Lee},\ and\ \citenamefont {Lu}}]{moon2019temporal}%
  \BibitemOpen
  \bibfield  {author} {\bibinfo {author} {\bibfnamefont {J.}~\bibnamefont
  {Moon}}, \bibinfo {author} {\bibfnamefont {W.}~\bibnamefont {Ma}}, \bibinfo
  {author} {\bibfnamefont {J.~H.}\ \bibnamefont {Shin}}, \bibinfo {author}
  {\bibfnamefont {F.}~\bibnamefont {Cai}}, \bibinfo {author} {\bibfnamefont
  {C.}~\bibnamefont {Du}}, \bibinfo {author} {\bibfnamefont {S.~H.}\
  \bibnamefont {Lee}},\ and\ \bibinfo {author} {\bibfnamefont {W.~D.}\
  \bibnamefont {Lu}},\ }\bibfield  {title} {\bibinfo {title} {Temporal data
  classification and forecasting using a memristor-based reservoir computing
  system},\ }\href@noop {} {\bibfield  {journal} {\bibinfo  {journal} {Nature
  Electronics}\ }\textbf {\bibinfo {volume} {2}},\ \bibinfo {pages} {480}
  (\bibinfo {year} {2019})}\BibitemShut {NoStop}%
\bibitem [{\citenamefont {Nakajima}\ \emph {et~al.}(2015)\citenamefont
  {Nakajima}, \citenamefont {Hauser}, \citenamefont {Li},\ and\ \citenamefont
  {Pfeifer}}]{nakajima2015information}%
  \BibitemOpen
  \bibfield  {author} {\bibinfo {author} {\bibfnamefont {K.}~\bibnamefont
  {Nakajima}}, \bibinfo {author} {\bibfnamefont {H.}~\bibnamefont {Hauser}},
  \bibinfo {author} {\bibfnamefont {T.}~\bibnamefont {Li}},\ and\ \bibinfo
  {author} {\bibfnamefont {R.}~\bibnamefont {Pfeifer}},\ }\bibfield  {title}
  {\bibinfo {title} {Information processing via physical soft body},\
  }\href@noop {} {\bibfield  {journal} {\bibinfo  {journal} {Scientific
  reports}\ }\textbf {\bibinfo {volume} {5}},\ \bibinfo {pages} {10487}
  (\bibinfo {year} {2015})}\BibitemShut {NoStop}%
\bibitem [{\citenamefont {Fernando}\ and\ \citenamefont
  {Sojakka}(2003)}]{fernando2003pattern}%
  \BibitemOpen
  \bibfield  {author} {\bibinfo {author} {\bibfnamefont {C.}~\bibnamefont
  {Fernando}}\ and\ \bibinfo {author} {\bibfnamefont {S.}~\bibnamefont
  {Sojakka}},\ }\bibfield  {title} {\bibinfo {title} {Pattern recognition in a
  bucket},\ }in\ \href@noop {} {\emph {\bibinfo {booktitle} {European
  conference on artificial life}}}\ (\bibinfo {organization} {Springer},\
  \bibinfo {year} {2003})\ pp.\ \bibinfo {pages} {588--597}\BibitemShut
  {NoStop}%
\bibitem [{\citenamefont {Dion}\ \emph {et~al.}(2018)\citenamefont {Dion},
  \citenamefont {Mejaouri},\ and\ \citenamefont
  {Sylvestre}}]{dion2018reservoir}%
  \BibitemOpen
  \bibfield  {author} {\bibinfo {author} {\bibfnamefont {G.}~\bibnamefont
  {Dion}}, \bibinfo {author} {\bibfnamefont {S.}~\bibnamefont {Mejaouri}},\
  and\ \bibinfo {author} {\bibfnamefont {J.}~\bibnamefont {Sylvestre}},\
  }\bibfield  {title} {\bibinfo {title} {Reservoir computing with a single
  delay-coupled non-linear mechanical oscillator},\ }\href@noop {} {\bibfield
  {journal} {\bibinfo  {journal} {Journal of Applied Physics}\ }\textbf
  {\bibinfo {volume} {124}},\ \bibinfo {pages} {152132} (\bibinfo {year}
  {2018})}\BibitemShut {NoStop}%
\bibitem [{\citenamefont {Nako}\ \emph {et~al.}(2020)\citenamefont {Nako},
  \citenamefont {Toprasertpong}, \citenamefont {Nakane}, \citenamefont {Wang},
  \citenamefont {Miyatake}, \citenamefont {Takenaka},\ and\ \citenamefont
  {Takagi}}]{nako2020proposal}%
  \BibitemOpen
  \bibfield  {author} {\bibinfo {author} {\bibfnamefont {E.}~\bibnamefont
  {Nako}}, \bibinfo {author} {\bibfnamefont {K.}~\bibnamefont {Toprasertpong}},
  \bibinfo {author} {\bibfnamefont {R.}~\bibnamefont {Nakane}}, \bibinfo
  {author} {\bibfnamefont {Z.}~\bibnamefont {Wang}}, \bibinfo {author}
  {\bibfnamefont {Y.}~\bibnamefont {Miyatake}}, \bibinfo {author}
  {\bibfnamefont {M.}~\bibnamefont {Takenaka}},\ and\ \bibinfo {author}
  {\bibfnamefont {S.}~\bibnamefont {Takagi}},\ }\bibfield  {title} {\bibinfo
  {title} {Proposal and experimental demonstration of reservoir computing using
  {H}f$_{0.5}${Z}r$_{0.5}${O}$_{2}$/{S}i {F}e{FET}s for neuromorphic
  applications},\ }in\ \href@noop {} {\emph {\bibinfo {booktitle} {2020 IEEE
  Symposium on VLSI Technology}}}\ (\bibinfo {organization} {IEEE},\ \bibinfo
  {year} {2020})\ pp.\ \bibinfo {pages} {1--2}\BibitemShut {NoStop}%
\bibitem [{\citenamefont {Torrejon}\ \emph {et~al.}(2017)\citenamefont
  {Torrejon}, \citenamefont {Riou}, \citenamefont {Araujo}, \citenamefont
  {Tsunegi}, \citenamefont {Khalsa}, \citenamefont {Querlioz}, \citenamefont
  {Bortolotti}, \citenamefont {Cros}, \citenamefont {Yakushiji}, \citenamefont
  {Fukushima} \emph {et~al.}}]{torrejon2017neuromorphic}%
  \BibitemOpen
  \bibfield  {author} {\bibinfo {author} {\bibfnamefont {J.}~\bibnamefont
  {Torrejon}}, \bibinfo {author} {\bibfnamefont {M.}~\bibnamefont {Riou}},
  \bibinfo {author} {\bibfnamefont {F.~A.}\ \bibnamefont {Araujo}}, \bibinfo
  {author} {\bibfnamefont {S.}~\bibnamefont {Tsunegi}}, \bibinfo {author}
  {\bibfnamefont {G.}~\bibnamefont {Khalsa}}, \bibinfo {author} {\bibfnamefont
  {D.}~\bibnamefont {Querlioz}}, \bibinfo {author} {\bibfnamefont
  {P.}~\bibnamefont {Bortolotti}}, \bibinfo {author} {\bibfnamefont
  {V.}~\bibnamefont {Cros}}, \bibinfo {author} {\bibfnamefont {K.}~\bibnamefont
  {Yakushiji}}, \bibinfo {author} {\bibfnamefont {A.}~\bibnamefont
  {Fukushima}}, \emph {et~al.},\ }\bibfield  {title} {\bibinfo {title}
  {Neuromorphic computing with nanoscale spintronic oscillators},\ }\href@noop
  {} {\bibfield  {journal} {\bibinfo  {journal} {Nature}\ }\textbf {\bibinfo
  {volume} {547}},\ \bibinfo {pages} {428} (\bibinfo {year}
  {2017})}\BibitemShut {NoStop}%
\bibitem [{\citenamefont {Kanao}\ \emph {et~al.}(2019)\citenamefont {Kanao},
  \citenamefont {Suto}, \citenamefont {Mizushima}, \citenamefont {Goto},
  \citenamefont {Tanamoto},\ and\ \citenamefont
  {Nagasawa}}]{kanao2019reservoir}%
  \BibitemOpen
  \bibfield  {author} {\bibinfo {author} {\bibfnamefont {T.}~\bibnamefont
  {Kanao}}, \bibinfo {author} {\bibfnamefont {H.}~\bibnamefont {Suto}},
  \bibinfo {author} {\bibfnamefont {K.}~\bibnamefont {Mizushima}}, \bibinfo
  {author} {\bibfnamefont {H.}~\bibnamefont {Goto}}, \bibinfo {author}
  {\bibfnamefont {T.}~\bibnamefont {Tanamoto}},\ and\ \bibinfo {author}
  {\bibfnamefont {T.}~\bibnamefont {Nagasawa}},\ }\bibfield  {title} {\bibinfo
  {title} {Reservoir computing on spin-torque oscillator array},\ }\href@noop
  {} {\bibfield  {journal} {\bibinfo  {journal} {Physical Review Applied}\
  }\textbf {\bibinfo {volume} {12}},\ \bibinfo {pages} {024052} (\bibinfo
  {year} {2019})}\BibitemShut {NoStop}%
\bibitem [{\citenamefont {Yamaguchi}\ \emph {et~al.}(2020)\citenamefont
  {Yamaguchi}, \citenamefont {Akashi}, \citenamefont {Tsunegi}, \citenamefont
  {Kubota}, \citenamefont {Nakajima},\ and\ \citenamefont
  {Taniguchi}}]{yamaguchi2020periodic}%
  \BibitemOpen
  \bibfield  {author} {\bibinfo {author} {\bibfnamefont {T.}~\bibnamefont
  {Yamaguchi}}, \bibinfo {author} {\bibfnamefont {N.}~\bibnamefont {Akashi}},
  \bibinfo {author} {\bibfnamefont {S.}~\bibnamefont {Tsunegi}}, \bibinfo
  {author} {\bibfnamefont {H.}~\bibnamefont {Kubota}}, \bibinfo {author}
  {\bibfnamefont {K.}~\bibnamefont {Nakajima}},\ and\ \bibinfo {author}
  {\bibfnamefont {T.}~\bibnamefont {Taniguchi}},\ }\bibfield  {title} {\bibinfo
  {title} {Periodic structure of memory function in spintronics reservoir with
  feedback current},\ }\href@noop {} {\bibfield  {journal} {\bibinfo  {journal}
  {Physical Review Research}\ }\textbf {\bibinfo {volume} {2}},\ \bibinfo
  {pages} {023389} (\bibinfo {year} {2020})}\BibitemShut {NoStop}%
\bibitem [{\citenamefont {Furuta}\ \emph {et~al.}(2018)\citenamefont {Furuta},
  \citenamefont {Fujii}, \citenamefont {Nakajima}, \citenamefont {Tsunegi},
  \citenamefont {Kubota}, \citenamefont {Suzuki},\ and\ \citenamefont
  {Miwa}}]{furuta2018macromagnetic}%
  \BibitemOpen
  \bibfield  {author} {\bibinfo {author} {\bibfnamefont {T.}~\bibnamefont
  {Furuta}}, \bibinfo {author} {\bibfnamefont {K.}~\bibnamefont {Fujii}},
  \bibinfo {author} {\bibfnamefont {K.}~\bibnamefont {Nakajima}}, \bibinfo
  {author} {\bibfnamefont {S.}~\bibnamefont {Tsunegi}}, \bibinfo {author}
  {\bibfnamefont {H.}~\bibnamefont {Kubota}}, \bibinfo {author} {\bibfnamefont
  {Y.}~\bibnamefont {Suzuki}},\ and\ \bibinfo {author} {\bibfnamefont
  {S.}~\bibnamefont {Miwa}},\ }\bibfield  {title} {\bibinfo {title}
  {Macromagnetic simulation for reservoir computing utilizing spin dynamics in
  magnetic tunnel junctions},\ }\href@noop {} {\bibfield  {journal} {\bibinfo
  {journal} {Physical Review Applied}\ }\textbf {\bibinfo {volume} {10}},\
  \bibinfo {pages} {034063} (\bibinfo {year} {2018})}\BibitemShut {NoStop}%
\bibitem [{\citenamefont {Nomura}\ \emph {et~al.}(2019)\citenamefont {Nomura},
  \citenamefont {Furuta}, \citenamefont {Tsujimoto}, \citenamefont
  {Kuwabiraki}, \citenamefont {Peper}, \citenamefont {Tamura}, \citenamefont
  {Miwa}, \citenamefont {Goto}, \citenamefont {Nakatani},\ and\ \citenamefont
  {Suzuki}}]{nomura2019reservoir}%
  \BibitemOpen
  \bibfield  {author} {\bibinfo {author} {\bibfnamefont {H.}~\bibnamefont
  {Nomura}}, \bibinfo {author} {\bibfnamefont {T.}~\bibnamefont {Furuta}},
  \bibinfo {author} {\bibfnamefont {K.}~\bibnamefont {Tsujimoto}}, \bibinfo
  {author} {\bibfnamefont {Y.}~\bibnamefont {Kuwabiraki}}, \bibinfo {author}
  {\bibfnamefont {F.}~\bibnamefont {Peper}}, \bibinfo {author} {\bibfnamefont
  {E.}~\bibnamefont {Tamura}}, \bibinfo {author} {\bibfnamefont
  {S.}~\bibnamefont {Miwa}}, \bibinfo {author} {\bibfnamefont {M.}~\bibnamefont
  {Goto}}, \bibinfo {author} {\bibfnamefont {R.}~\bibnamefont {Nakatani}},\
  and\ \bibinfo {author} {\bibfnamefont {Y.}~\bibnamefont {Suzuki}},\
  }\bibfield  {title} {\bibinfo {title} {Reservoir computing with
  dipole-coupled nanomagnets},\ }\href@noop {} {\bibfield  {journal} {\bibinfo
  {journal} {Japanese Journal of Applied Physics}\ }\textbf {\bibinfo {volume}
  {58}},\ \bibinfo {pages} {070901} (\bibinfo {year} {2019})}\BibitemShut
  {NoStop}%
\bibitem [{\citenamefont {Nakane}\ \emph {et~al.}(2018)\citenamefont {Nakane},
  \citenamefont {Tanaka},\ and\ \citenamefont {Hirose}}]{nakane2018reservoir}%
  \BibitemOpen
  \bibfield  {author} {\bibinfo {author} {\bibfnamefont {R.}~\bibnamefont
  {Nakane}}, \bibinfo {author} {\bibfnamefont {G.}~\bibnamefont {Tanaka}},\
  and\ \bibinfo {author} {\bibfnamefont {A.}~\bibnamefont {Hirose}},\
  }\bibfield  {title} {\bibinfo {title} {Reservoir computing with spin waves
  excited in a garnet film},\ }\href@noop {} {\bibfield  {journal} {\bibinfo
  {journal} {IEEE Access}\ }\textbf {\bibinfo {volume} {6}},\ \bibinfo {pages}
  {4462} (\bibinfo {year} {2018})}\BibitemShut {NoStop}%
\bibitem [{\citenamefont {Nakane}\ \emph {et~al.}(2019)\citenamefont {Nakane},
  \citenamefont {Tanaka},\ and\ \citenamefont {Hirose}}]{nakane2019spin}%
  \BibitemOpen
  \bibfield  {author} {\bibinfo {author} {\bibfnamefont {R.}~\bibnamefont
  {Nakane}}, \bibinfo {author} {\bibfnamefont {G.}~\bibnamefont {Tanaka}},\
  and\ \bibinfo {author} {\bibfnamefont {A.}~\bibnamefont {Hirose}},\
  }\bibfield  {title} {\bibinfo {title} {In a spin-wave reservoir for machine
  learning},\ }in\ \href@noop {} {\emph {\bibinfo {booktitle} {2019
  International Joint Conference on Neural Networks (IJCNN)}}}\ (\bibinfo
  {organization} {IEEE},\ \bibinfo {year} {2019})\ pp.\ \bibinfo {pages}
  {1--9}\BibitemShut {NoStop}%
\bibitem [{\citenamefont {Mahmoud}\ \emph {et~al.}(2020)\citenamefont
  {Mahmoud}, \citenamefont {Ciubotaru}, \citenamefont {Vanderveken},
  \citenamefont {Chumak}, \citenamefont {Hamdioui}, \citenamefont {Adelmann},\
  and\ \citenamefont {Cotofana}}]{mahmoud2020introduction}%
  \BibitemOpen
  \bibfield  {author} {\bibinfo {author} {\bibfnamefont {A.}~\bibnamefont
  {Mahmoud}}, \bibinfo {author} {\bibfnamefont {F.}~\bibnamefont {Ciubotaru}},
  \bibinfo {author} {\bibfnamefont {F.}~\bibnamefont {Vanderveken}}, \bibinfo
  {author} {\bibfnamefont {A.~V.}\ \bibnamefont {Chumak}}, \bibinfo {author}
  {\bibfnamefont {S.}~\bibnamefont {Hamdioui}}, \bibinfo {author}
  {\bibfnamefont {C.}~\bibnamefont {Adelmann}},\ and\ \bibinfo {author}
  {\bibfnamefont {S.}~\bibnamefont {Cotofana}},\ }\bibfield  {title} {\bibinfo
  {title} {Introduction to spin wave computing},\ }\href@noop {} {\bibfield
  {journal} {\bibinfo  {journal} {Journal of Applied Physics}\ }\textbf
  {\bibinfo {volume} {128}},\ \bibinfo {pages} {161101} (\bibinfo {year}
  {2020})}\BibitemShut {NoStop}%
\bibitem [{\citenamefont {Luko{\v{s}}evi{\v{c}}ius}\ and\ \citenamefont
  {Jaeger}(2009)}]{lukovsevivcius2009reservoir}%
  \BibitemOpen
  \bibfield  {author} {\bibinfo {author} {\bibfnamefont {M.}~\bibnamefont
  {Luko{\v{s}}evi{\v{c}}ius}}\ and\ \bibinfo {author} {\bibfnamefont
  {H.}~\bibnamefont {Jaeger}},\ }\bibfield  {title} {\bibinfo {title}
  {Reservoir computing approaches to recurrent neural network training},\
  }\href@noop {} {\bibfield  {journal} {\bibinfo  {journal} {Computer Science
  Review}\ }\textbf {\bibinfo {volume} {3}},\ \bibinfo {pages} {127} (\bibinfo
  {year} {2009})}\BibitemShut {NoStop}%
\bibitem [{\citenamefont {Rana}\ and\ \citenamefont
  {Otani}(2019)}]{rana2019towards}%
  \BibitemOpen
  \bibfield  {author} {\bibinfo {author} {\bibfnamefont {B.}~\bibnamefont
  {Rana}}\ and\ \bibinfo {author} {\bibfnamefont {Y.}~\bibnamefont {Otani}},\
  }\bibfield  {title} {\bibinfo {title} {Towards magnonic devices based on
  voltage-controlled magnetic anisotropy},\ }\href@noop {} {\bibfield
  {journal} {\bibinfo  {journal} {Communications Physics}\ }\textbf {\bibinfo
  {volume} {2}},\ \bibinfo {pages} {1} (\bibinfo {year} {2019})}\BibitemShut
  {NoStop}%
\bibitem [{\citenamefont {Hubert}\ and\ \citenamefont
  {Sch{\"a}fer}(2008)}]{hubert2008magnetic}%
  \BibitemOpen
  \bibfield  {author} {\bibinfo {author} {\bibfnamefont {A.}~\bibnamefont
  {Hubert}}\ and\ \bibinfo {author} {\bibfnamefont {R.}~\bibnamefont
  {Sch{\"a}fer}},\ }\href@noop {} {\emph {\bibinfo {title} {Magnetic domains:
  the analysis of magnetic microstructures}}}\ (\bibinfo  {publisher} {Springer
  Science \& Business Media},\ \bibinfo {year} {2008})\ Chap.\ \bibinfo
  {chapter} {5.6.1}\BibitemShut {NoStop}%
\bibitem [{\citenamefont {Stancil}\ and\ \citenamefont
  {Prabhakar}(2009)}]{stancil2009spin}%
  \BibitemOpen
  \bibfield  {author} {\bibinfo {author} {\bibfnamefont {D.~D.}\ \bibnamefont
  {Stancil}}\ and\ \bibinfo {author} {\bibfnamefont {A.}~\bibnamefont
  {Prabhakar}},\ }\bibinfo {title} {Spin waves}\ (\bibinfo  {publisher}
  {Springer},\ \bibinfo {year} {2009})\ Chap.~\bibinfo {chapter}
  {9}\BibitemShut {NoStop}%
\bibitem [{\citenamefont {Craik}\ and\ \citenamefont
  {Cooper}(1978)}]{craik1978bias}%
  \BibitemOpen
  \bibfield  {author} {\bibinfo {author} {\bibfnamefont {D.}~\bibnamefont
  {Craik}}\ and\ \bibinfo {author} {\bibfnamefont {P.}~\bibnamefont {Cooper}},\
  }\bibfield  {title} {\bibinfo {title} {Bias magnet design for bubble memory
  devices},\ }\href@noop {} {\bibfield  {journal} {\bibinfo  {journal} {IEEE
  Transactions on Magnetics}\ }\textbf {\bibinfo {volume} {14}},\ \bibinfo
  {pages} {306} (\bibinfo {year} {1978})}\BibitemShut {NoStop}%
\bibitem [{\citenamefont {Vansteenkiste}\ \emph {et~al.}(2014)\citenamefont
  {Vansteenkiste}, \citenamefont {Leliaert}, \citenamefont {Dvornik},
  \citenamefont {Helsen}, \citenamefont {Garcia-Sanchez},\ and\ \citenamefont
  {Van~Waeyenberge}}]{vansteenkiste2014design}%
  \BibitemOpen
  \bibfield  {author} {\bibinfo {author} {\bibfnamefont {A.}~\bibnamefont
  {Vansteenkiste}}, \bibinfo {author} {\bibfnamefont {J.}~\bibnamefont
  {Leliaert}}, \bibinfo {author} {\bibfnamefont {M.}~\bibnamefont {Dvornik}},
  \bibinfo {author} {\bibfnamefont {M.}~\bibnamefont {Helsen}}, \bibinfo
  {author} {\bibfnamefont {F.}~\bibnamefont {Garcia-Sanchez}},\ and\ \bibinfo
  {author} {\bibfnamefont {B.}~\bibnamefont {Van~Waeyenberge}},\ }\bibfield
  {title} {\bibinfo {title} {The design and verification of mumax3},\
  }\href@noop {} {\bibfield  {journal} {\bibinfo  {journal} {AIP advances}\
  }\textbf {\bibinfo {volume} {4}},\ \bibinfo {pages} {107133} (\bibinfo {year}
  {2014})}\BibitemShut {NoStop}%
\bibitem [{\citenamefont {Anderson}(1964)}]{anderson1964molecular}%
  \BibitemOpen
  \bibfield  {author} {\bibinfo {author} {\bibfnamefont {E.~E.}\ \bibnamefont
  {Anderson}},\ }\bibfield  {title} {\bibinfo {title} {Molecular field model
  and the magnetization of yig},\ }\href@noop {} {\bibfield  {journal}
  {\bibinfo  {journal} {Physical Review}\ }\textbf {\bibinfo {volume} {134}},\
  \bibinfo {pages} {A1581} (\bibinfo {year} {1964})}\BibitemShut {NoStop}%
\bibitem [{\citenamefont {Klingler}\ \emph {et~al.}(2014)\citenamefont
  {Klingler}, \citenamefont {Chumak}, \citenamefont {Mewes}, \citenamefont
  {Khodadadi}, \citenamefont {Mewes}, \citenamefont {Dubs}, \citenamefont
  {Surzhenko}, \citenamefont {Hillebrands},\ and\ \citenamefont
  {Conca}}]{klingler2014measurements}%
  \BibitemOpen
  \bibfield  {author} {\bibinfo {author} {\bibfnamefont {S.}~\bibnamefont
  {Klingler}}, \bibinfo {author} {\bibfnamefont {A.~V.}\ \bibnamefont
  {Chumak}}, \bibinfo {author} {\bibfnamefont {T.}~\bibnamefont {Mewes}},
  \bibinfo {author} {\bibfnamefont {B.}~\bibnamefont {Khodadadi}}, \bibinfo
  {author} {\bibfnamefont {C.}~\bibnamefont {Mewes}}, \bibinfo {author}
  {\bibfnamefont {C.}~\bibnamefont {Dubs}}, \bibinfo {author} {\bibfnamefont
  {O.}~\bibnamefont {Surzhenko}}, \bibinfo {author} {\bibfnamefont
  {B.}~\bibnamefont {Hillebrands}},\ and\ \bibinfo {author} {\bibfnamefont
  {A.}~\bibnamefont {Conca}},\ }\bibfield  {title} {\bibinfo {title}
  {Measurements of the exchange stiffness of {YIG} films using broadband
  ferromagnetic resonance techniques},\ }\href@noop {} {\bibfield  {journal}
  {\bibinfo  {journal} {Journal of Physics D: Applied Physics}\ }\textbf
  {\bibinfo {volume} {48}},\ \bibinfo {pages} {015001} (\bibinfo {year}
  {2014})}\BibitemShut {NoStop}%
\bibitem [{\citenamefont {Manuilov}\ \emph {et~al.}(2009)\citenamefont
  {Manuilov}, \citenamefont {Khartsev},\ and\ \citenamefont
  {Grishin}}]{manuilov2009pulsed}%
  \BibitemOpen
  \bibfield  {author} {\bibinfo {author} {\bibfnamefont {S.~A.}\ \bibnamefont
  {Manuilov}}, \bibinfo {author} {\bibfnamefont {S.}~\bibnamefont {Khartsev}},\
  and\ \bibinfo {author} {\bibfnamefont {A.~M.}\ \bibnamefont {Grishin}},\
  }\bibfield  {title} {\bibinfo {title} {Pulsed laser deposited
  {Y}$_{3}${F}e$_{5}${O}$_{12}$ films: Nature of magnetic anisotropy {I}},\
  }\href@noop {} {\bibfield  {journal} {\bibinfo  {journal} {Journal of Applied
  Physics}\ }\textbf {\bibinfo {volume} {106}},\ \bibinfo {pages} {123917}
  (\bibinfo {year} {2009})}\BibitemShut {NoStop}%
\bibitem [{\citenamefont {Dubs}\ \emph {et~al.}(2020)\citenamefont {Dubs},
  \citenamefont {Surzhenko}, \citenamefont {Thomas}, \citenamefont {Osten},
  \citenamefont {Schneider}, \citenamefont {Lenz}, \citenamefont {Grenzer},
  \citenamefont {H{\"u}bner},\ and\ \citenamefont {Wendler}}]{dubs2020low}%
  \BibitemOpen
  \bibfield  {author} {\bibinfo {author} {\bibfnamefont {C.}~\bibnamefont
  {Dubs}}, \bibinfo {author} {\bibfnamefont {O.}~\bibnamefont {Surzhenko}},
  \bibinfo {author} {\bibfnamefont {R.}~\bibnamefont {Thomas}}, \bibinfo
  {author} {\bibfnamefont {J.}~\bibnamefont {Osten}}, \bibinfo {author}
  {\bibfnamefont {T.}~\bibnamefont {Schneider}}, \bibinfo {author}
  {\bibfnamefont {K.}~\bibnamefont {Lenz}}, \bibinfo {author} {\bibfnamefont
  {J.}~\bibnamefont {Grenzer}}, \bibinfo {author} {\bibfnamefont
  {R.}~\bibnamefont {H{\"u}bner}},\ and\ \bibinfo {author} {\bibfnamefont
  {E.}~\bibnamefont {Wendler}},\ }\bibfield  {title} {\bibinfo {title} {Low
  damping and microstructural perfection of sub-40nm-thin yttrium iron garnet
  films grown by liquid phase epitaxy},\ }\href@noop {} {\bibfield  {journal}
  {\bibinfo  {journal} {Physical Review Materials}\ }\textbf {\bibinfo {volume}
  {4}},\ \bibinfo {pages} {024416} (\bibinfo {year} {2020})}\BibitemShut
  {NoStop}%
\bibitem [{\citenamefont {Pirro}\ \emph {et~al.}(2014)\citenamefont {Pirro},
  \citenamefont {Br{\"a}cher}, \citenamefont {Chumak}, \citenamefont
  {L{\"a}gel}, \citenamefont {Dubs}, \citenamefont {Surzhenko}, \citenamefont
  {G{\"o}rnert}, \citenamefont {Leven},\ and\ \citenamefont
  {Hillebrands}}]{pirro2014spin}%
  \BibitemOpen
  \bibfield  {author} {\bibinfo {author} {\bibfnamefont {P.}~\bibnamefont
  {Pirro}}, \bibinfo {author} {\bibfnamefont {T.}~\bibnamefont {Br{\"a}cher}},
  \bibinfo {author} {\bibfnamefont {A.}~\bibnamefont {Chumak}}, \bibinfo
  {author} {\bibfnamefont {B.}~\bibnamefont {L{\"a}gel}}, \bibinfo {author}
  {\bibfnamefont {C.}~\bibnamefont {Dubs}}, \bibinfo {author} {\bibfnamefont
  {O.}~\bibnamefont {Surzhenko}}, \bibinfo {author} {\bibfnamefont
  {P.}~\bibnamefont {G{\"o}rnert}}, \bibinfo {author} {\bibfnamefont
  {B.}~\bibnamefont {Leven}},\ and\ \bibinfo {author} {\bibfnamefont
  {B.}~\bibnamefont {Hillebrands}},\ }\bibfield  {title} {\bibinfo {title}
  {Spin-wave excitation and propagation in microstructured waveguides of
  yttrium iron garnet/{P}t bilayers},\ }\href@noop {} {\bibfield  {journal}
  {\bibinfo  {journal} {Applied Physics Letters}\ }\textbf {\bibinfo {volume}
  {104}},\ \bibinfo {pages} {012402} (\bibinfo {year} {2014})}\BibitemShut
  {NoStop}%
\bibitem [{\citenamefont {Ding}\ \emph {et~al.}(2020)\citenamefont {Ding},
  \citenamefont {Liu}, \citenamefont {Chang},\ and\ \citenamefont
  {Wu}}]{ding2020sputtering}%
  \BibitemOpen
  \bibfield  {author} {\bibinfo {author} {\bibfnamefont {J.}~\bibnamefont
  {Ding}}, \bibinfo {author} {\bibfnamefont {T.}~\bibnamefont {Liu}}, \bibinfo
  {author} {\bibfnamefont {H.}~\bibnamefont {Chang}},\ and\ \bibinfo {author}
  {\bibfnamefont {M.}~\bibnamefont {Wu}},\ }\bibfield  {title} {\bibinfo
  {title} {Sputtering growth of low-damping yttrium-iron-garnet thin films},\
  }\href@noop {} {\bibfield  {journal} {\bibinfo  {journal} {IEEE Magnetics
  Letters}\ }\textbf {\bibinfo {volume} {11}},\ \bibinfo {pages} {1} (\bibinfo
  {year} {2020})}\BibitemShut {NoStop}%
\bibitem [{\citenamefont {Schmidt}\ \emph {et~al.}(2020)\citenamefont
  {Schmidt}, \citenamefont {Hauser}, \citenamefont {Trempler}, \citenamefont
  {Paleschke},\ and\ \citenamefont {Papaioannou}}]{schmidt2020ultra}%
  \BibitemOpen
  \bibfield  {author} {\bibinfo {author} {\bibfnamefont {G.}~\bibnamefont
  {Schmidt}}, \bibinfo {author} {\bibfnamefont {C.}~\bibnamefont {Hauser}},
  \bibinfo {author} {\bibfnamefont {P.}~\bibnamefont {Trempler}}, \bibinfo
  {author} {\bibfnamefont {M.}~\bibnamefont {Paleschke}},\ and\ \bibinfo
  {author} {\bibfnamefont {E.~T.}\ \bibnamefont {Papaioannou}},\ }\bibfield
  {title} {\bibinfo {title} {Ultra thin films of yttrium iron garnet with very
  low damping: A review},\ }\href@noop {} {\bibfield  {journal} {\bibinfo
  {journal} {physica status solidi (b)}\ }\textbf {\bibinfo {volume} {257}},\
  \bibinfo {pages} {1900644} (\bibinfo {year} {2020})}\BibitemShut {NoStop}%
\bibitem [{\citenamefont {Testa}\ \emph {et~al.}(2007)\citenamefont {Testa},
  \citenamefont {Akram}, \citenamefont {Burch}, \citenamefont {Carpinelli},
  \citenamefont {Chang}, \citenamefont {Dinavahi}, \citenamefont {Hatziadoniu},
  \citenamefont {Grady}, \citenamefont {Gunther}, \citenamefont {Halpin} \emph
  {et~al.}}]{testa2007interharmonics}%
  \BibitemOpen
  \bibfield  {author} {\bibinfo {author} {\bibfnamefont {A.}~\bibnamefont
  {Testa}}, \bibinfo {author} {\bibfnamefont {M.}~\bibnamefont {Akram}},
  \bibinfo {author} {\bibfnamefont {R.}~\bibnamefont {Burch}}, \bibinfo
  {author} {\bibfnamefont {G.}~\bibnamefont {Carpinelli}}, \bibinfo {author}
  {\bibfnamefont {G.}~\bibnamefont {Chang}}, \bibinfo {author} {\bibfnamefont
  {V.}~\bibnamefont {Dinavahi}}, \bibinfo {author} {\bibfnamefont
  {C.}~\bibnamefont {Hatziadoniu}}, \bibinfo {author} {\bibfnamefont
  {W.}~\bibnamefont {Grady}}, \bibinfo {author} {\bibfnamefont
  {E.}~\bibnamefont {Gunther}}, \bibinfo {author} {\bibfnamefont
  {M.}~\bibnamefont {Halpin}}, \emph {et~al.},\ }\bibfield  {title} {\bibinfo
  {title} {Interharmonics: Theory and modeling},\ }\href@noop {} {\bibfield
  {journal} {\bibinfo  {journal} {IEEE Transactions on Power Delivery}\
  }\textbf {\bibinfo {volume} {22}},\ \bibinfo {pages} {2335} (\bibinfo {year}
  {2007})}\BibitemShut {NoStop}%
\bibitem [{\citenamefont {Jaeger}(2002)}]{jaeger2002short}%
  \BibitemOpen
  \bibfield  {author} {\bibinfo {author} {\bibfnamefont {H.}~\bibnamefont
  {Jaeger}},\ }\bibfield  {title} {\bibinfo {title} {Short term memory in echo
  state networks},\ }\href@noop {} {\bibfield  {journal} {\bibinfo  {journal}
  {GMD-German National Research Institute for Computer Science GMD Report}\
  }\textbf {\bibinfo {volume} {152}} (\bibinfo {year} {2002})}\BibitemShut
  {NoStop}%
\bibitem [{\citenamefont {Atiya}\ and\ \citenamefont
  {Parlos}(2000)}]{atiya2000new}%
  \BibitemOpen
  \bibfield  {author} {\bibinfo {author} {\bibfnamefont {A.~F.}\ \bibnamefont
  {Atiya}}\ and\ \bibinfo {author} {\bibfnamefont {A.~G.}\ \bibnamefont
  {Parlos}},\ }\bibfield  {title} {\bibinfo {title} {New results on recurrent
  network training: unifying the algorithms and accelerating convergence},\
  }\href@noop {} {\bibfield  {journal} {\bibinfo  {journal} {IEEE transactions
  on neural networks}\ }\textbf {\bibinfo {volume} {11}},\ \bibinfo {pages}
  {697} (\bibinfo {year} {2000})}\BibitemShut {NoStop}%
\bibitem [{\citenamefont {Appeltant}(2012)}]{appeltant2012reservoir}%
  \BibitemOpen
  \bibfield  {author} {\bibinfo {author} {\bibfnamefont {L.}~\bibnamefont
  {Appeltant}},\ }\bibinfo {title} {Reservoir computing based on
  delay-dynamical systems}\ (\bibinfo  {publisher} {These de Doctorat, Vrije
  Universiteit Brussel/Universitat de les Illes Balears},\ \bibinfo {year}
  {2012})\ Chap.\ \bibinfo {chapter} {5.2}\BibitemShut {NoStop}%
\bibitem [{\citenamefont {Inubushi}\ and\ \citenamefont
  {Yoshimura}(2017)}]{inubushi2017reservoir}%
  \BibitemOpen
  \bibfield  {author} {\bibinfo {author} {\bibfnamefont {M.}~\bibnamefont
  {Inubushi}}\ and\ \bibinfo {author} {\bibfnamefont {K.}~\bibnamefont
  {Yoshimura}},\ }\bibfield  {title} {\bibinfo {title} {Reservoir computing
  beyond memory-nonlinearity trade-off},\ }\href@noop {} {\bibfield  {journal}
  {\bibinfo  {journal} {Scientific reports}\ }\textbf {\bibinfo {volume} {7}},\
  \bibinfo {pages} {1} (\bibinfo {year} {2017})}\BibitemShut {NoStop}%
\end{thebibliography}
\providecommand{\noopsort}[1]{}\providecommand{\singleletter}[1]{#1}%

\end{document}